\pdfoutput=1
\documentclass[onecolumn,amsmath,amssymb,showpacs]{revtex4-1}
\usepackage{graphicx,color,xcolor,hyperref,amssymb,textcomp}
\usepackage{graphicx,color,xcolor,hyperref,amssymb,textcomp}
\definecolor{dgreen}{rgb}{0,0.7,0}

\begin{document}

\def\be{\begin{equation}}
\def\ee{\end{equation}}
\def\bea{\begin{eqnarray}}
\def\eea{\end{eqnarray}}
\def\l{\label}

\newcommand{\eref}[1]{Eq.~(\ref{#1})}%
\newcommand{\Eref}[1]{Equation~(\ref{#1})}%
\newcommand{\fref}[1]{Fig.~\ref{#1}} %
\newcommand{\Fref}[1]{Figure~\ref{#1}}%
\newcommand{\sref}[1]{Sec.~\ref{#1}}%
\newcommand{\Sref}[1]{Section~\ref{#1}}%
\newcommand{\aref}[1]{Appendix~\ref{#1}}%
\newcommand{\sgn}[1]{\mathrm{sgn}({#1})}%
\newcommand{\erfc}{\mathrm{erfc}}%
\newcommand{\erf}{\mathrm{erf}}%
%%%%%%%%%%%%%%%%%%%%%%%%%%%%%%%%%%%%%%%%%%%%%%%%%%%%%%%

\title{\mbox{Local time of diffusion with stochastic resetting}}

\author{{\normalsize{}Arnab Pal$^{1}$, Rakesh Chatterjee$^{2}$, Shlomi Reuveni$^{1}$, \& Anupam Kundu$^{3}$}
{\normalsize{}}}

\affiliation{\noindent \textit{$^{1}$School of Chemistry, The Center for Physics and Chemistry of Living Systems, The Raymond and Beverly Sackler Center for Computational Molecular and Materials Science,\\ \& The Mark Ratner Institute for Single Molecule Chemistry, Tel Aviv University, Tel Aviv 6997801, Israel}}

\affiliation{\noindent \textit{$^{2}$School of Mechanical Engineering, The Center for Physics and Chemistry of Living Systems \\ \& The Raymond and Beverly Sackler Center for Computational Molecular and Materials Science, Tel Aviv University, Tel Aviv 6997801, Israel}}

\affiliation{\noindent \textit{$^{3}$International Centre for Theoretical Sciences, TIFR, Bangalore 560012, India}}

\begin{abstract}

Diffusion with stochastic resetting has recently emerged as a powerful modeling tool with a myriad of potential applications. Here, we study local time in this model,
covering situations of free and biased diffusion with, and without, the presence of an absorbing boundary. Given a Brownian trajectory that evolved for $t$ units of
time, the local time is simply defined as the total time the trajectory spent in a small vicinity of its initial position. However, as Brownian trajectories
are stochastic --- the local time itself is a random variable which fluctuates round and about its mean value. In the past, the statistics of these fluctuations
has been quantified in detail; but not in the presence of resetting which biases the particle to spend more time near its starting point. Here, we extend past results to include
the possibility of stochastic resetting with, and without, the presence of an absorbing boundary and/or drift. We obtain exact results for the moments and distribution
of the local time and these reveal that its statistics usually admits a simple form in the long-time limit. And yet, while fluctuations in the absence of
stochastic resetting are typically non-Gaussian --- resetting gives rise to Gaussian  fluctuations. The analytical findings presented herein are in excellent agreement with numerical simulations.

\end{abstract}
\date{\today}
%\pacs{}
\maketitle

\section{Introduction}
\l{intro}
\noindent

%%%%%%%%%%%%%%%%%%%%%%%%%%%%%%%%%%%%
\begin{figure}[b!]
\includegraphics[width=13cm,height=5.7cm]{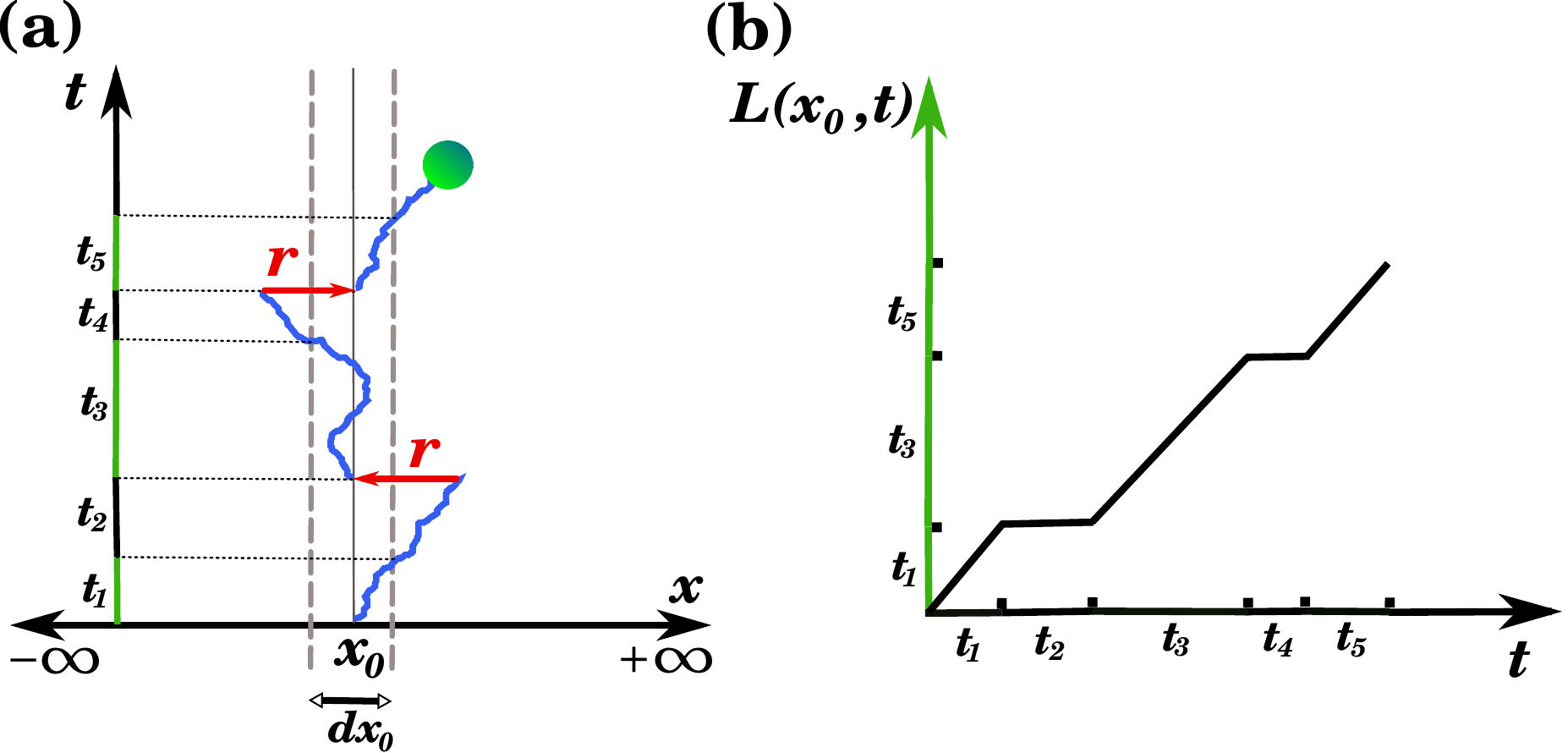}
\caption{Local time of diffusion with stochastic resetting: (a) Schematic diagram showing a sample trajectory of a Brownian particle diffusing in an infinite domain under stochastic resetting with rate $r$. Time intervals in which the particle is found within an infinitesimal vicinity of its starting position $x_0$, were denoted $\{t_1,t_3,t_5\}$ and highlighted in green; (b) The local time $L(x_0,t)$ that will be studied here is the total time the particle spent in the vicinity of the starting point $x_0$ during an observation time $t$. The time evolution of $L(x_0,t)$ is plotted for the trajectory in (a).}
\l{fig:localtime}
\end{figure}
%%%%%%%%%%%%%%%%%%%%%%%%%%%%%%%%%%%%

The local time of a particle is defined as the total time it spends in the vicinity of a distinguished point in space \cite{Levy,Knight,Feller,Satya-current-science}. 
As such, the local time provides interesting information with regard to the spatio-temporal properties of the particle's trajectory and is thus a useful measure with various applications. The local time can e.g., be used to quantify the time a molecule spends in the vicinity of a region in space where it is more likely to undergo a chemical reaction.  
Knowledge of the local time and its statistics can thus help improve our understanding of chemical reactions and catalytic processes \cite{Agmon1,Agmon2,porousD,reactive,bacteria}. In the past, the local time has been studied in various different scenarios e.g., diffusion in bounded domains \cite{RBM}, diffusion in a potential landscape \cite{Majumdar-Comtet2002}, diffusion in a random potential landscape \cite{Sabhapandit:2006}, and diffusion on a graph \cite{graph}. The local time has also been studied for random processes other than simple diffusion e.g., uniform empirical process \cite{Yor} and Brownian excursions \cite{Louchard}. However, in all cases mentioned above authors considered time evolution in the absence of stochastic resetting which recently attracted considerable attention \cite{Restart1, Restart2,Restart3,Restart4,Restart5,KM,
restart_conc1, restart_conc2,restart_conc3,restart_conc4,restart_conc5,restart_conc6,restart_conc7,restart_conc8, restart_conc10,restart_conc11,restart_conc16,restart_conc17,Hugo-Satya,restart_conc18,restart_conc21,underdamped,Satya-refractory,Masoliver-telegraphic,LukaszCTRW}. Consequently, very little is known about the local time of stochastic processes that undergo restart.

Restart is an emerging and overarching topic in physics which is currently being studied in relation to e.g., first-passage and search theory \cite{
Benichou-review,
Restart-Search1,Restart-Search2,Restart-Search3,Chechkin,restart_conc19,restart_conc20,ReuveniPRL,PalReuveniPRL,branching,drift-diffusion},
stochastic thermodynamics \cite{restart_thermo1,restart_thermo2}, optimization theory \cite{restart_conc10,Optimization}, quantum mechanics \cite{Quantum1,Quantum2}, and in connection with multiple different problems in statistical and biological physics \cite{Restart-Biophysics1,Restart-Biophysics2, Restart-Biophysics3, Restart-Biophysics4,Restart-Biophysics5,Restart-Biophysics6}. Among different models, a place of honor is reserved to diffusion with stochastic resetting as this played a central role in shaping our understanding of restart phenomena \cite{Restart1}. In diffusion with stochastic resetting, one considers a Brownian particle that is returned to its initial position at random time epochs. The system then exhibits non-trivial transient dynamics: as time progresses, an inner core region near the resetting point relaxes to a steady state while an outer region continues to be transient. The location of the boundary separating the two regions moves away from the origin as a power law in  time \cite{restart_conc5}. At very long times, the system will  eventually reach a non-equilibrium steady state, but fluctuations there were found to be highly non-Gaussian \cite{Restart1,Restart2}. Many other properties of diffusion with stochastic resetting (dynamical, first-passage, etc.) have also been studied extensively. In comparison, much less is known about observables (functionals) of Brownian trajectories with stochastic resetting as these are just starting to be studied. 

In a recent paper \cite{restart_conc17}, the authors examined large deviations 
of time-additive functions or observables of Markov processes with resetting and showed how this could also be applied to Brownian motion as a special case. Using renewal properties of the resetting dynamics, they derived a formula which links generating functions with and without resetting; and then used this formula to, e.g., compute the large deviations in the area covered by an Ornstein-Uhlenbeck trajectory with resetting. In a followup study \cite{Hugo-Satya}, the authors focused on additive functionals of Brownian motion with resetting. They derived a large deviation principle in the presence of resetting and identified the large deviation rate function in terms of a variational formula involving the large deviation rate functions without resetting. The power of this approach was demonstrated on three different Brownian functionals: the positive occupation time, the area, and the absolute area. Here, we study a different time additive functional of Brownian motion with resetting: the local time
(\fref{fig:localtime}).

Stochastic trajectories that are governed by resetting dynamics share a basic feature. By definition, upon resetting the particle returns to its initial position (Fig. \ref{fig:localtime}a). 
Thus, one observes a  probability current directed at the resetting location which naturally motivates us to study the local time of the particle near this attractor. 
Local time is a quantitative measure of the total time spent by a particle at an infinitesimal vicinity of any arbitrary location \cite{Majumdar-Comtet2002,Sabhapandit:2006,Satya-current-science}
and here we will be interested in the statistical properties of this observable (moments and distribution) in the presence of resetting.
In principle, these can always be determined by considering all the possible trajectories that emanate from $x_0$ and reset back to this point at a constant rate $r$. One successful method that allows
one to do this in practice is due to Feynman and Kac \cite{Satya-current-science,Schehr-review,Feynman-Kac} and we will employ this methodology extensively in this paper.

The remainder of this paper is structured as follows. In \sref{reset-dynamics}, we review the model of diffusion with stochastic resetting and describe the setup we consider. In \sref{methods}, we provide an outline of the general Feynman-Kac formalism which is used to compute local time statistics. In \sref{simple}, we compute the local time statistics of simple diffusion with, and without, stochastic resetting. Two case studies are analyzed: (i) unrestricted diffusion in 1D; and (ii) diffusion on the semi-infinite line with an absorbing boundary at the origin. Results in this section extend those that were previously reported in  \cite{Sabhapandit:2006}. In \sref{drift-diffusion}, we compute the local time statistics of drift-diffusion with, and without, stochastic resetting. Thus, we consider a Brownian particle with a constant drift velocity $\lambda$. Once again, two different case studies are analyzed: unrestricted diffusion and diffusion with an absorbing boundary at the origin. In both \sref{simple} and \sref{drift-diffusion} alike, we give exact results for the moments and distribution function of the local time. Asymptotic, long-time analysis of the local time and its statistics is also provided. \sref{conclusion} concludes this work. For clarity of the presentation some details have been reserved to the appendix.

%%%%%%%%%%%%%%%%%%%%%%%%%%%%%%%%%%%%%%%%%%%%%%%%%%%%%%%
\section{The model}
\l{reset-dynamics}
\noindent
We consider a single Brownian particle undergoing diffusion in one dimension in the presence of an external potential $V(x)$. The 
motion of the particle is described by the overdamped Langevin equation
\bea
\gamma \frac{dx}{d\tau}=F(x)+\eta(\tau)~,~~~\text{with}~~F(x)=-V'(x)~,
\l{eq:eom}
\eea
where the prime denotes differentiation with respect to the space variable $x$. Here $\gamma$ is the damping coefficient and $\eta(\tau)$ is a Gaussian white noise with the following statistical properties
$\langle\eta(\tau)\rangle=0$ and
$\langle\eta(\tau)\eta(\tau')\rangle=2D\delta(\tau-\tau')$ where $D$ is the diffusion coefficient. In what follows, we take $\gamma$ to be unity for simplicity. 
In addition to the above dynamics, the particle is intermittently stopped and then immediately brought to a fixed position $x_r$. In this paper, we take this point to be identical to the initial position of the particle. Thus, after each such resetting event the particle starts its motion from $x_0$. This procedure of resetting the particle's position is performed at a constant rate $r$. 
The dynamics of the particle then reads
\begin{equation}
\begin{array}{l}
x(t+dt)=\left\{ \begin{array}{lll}
x_0 ~~~~~~~~~~~~~~~~~~~~~~~&  & \text{with~probability~ }rdt\text{ }\\
 & \text{ \ \ }\\
x(t)+F(x)dt+\eta dt &  & \text{with~probability~ }1-rdt~.
\end{array}\right.\text{ }\end{array}
\label{eom}
\end{equation}
A schematic cartoon of a trajectory with resetting was shown in Fig.~\ref{fig:localtime}a.
In the following sections, we compute the probability distribution of the local time for two different cases: (i) $F(x)=0$ and (ii) $F(x)=\lambda$, where $\lambda$ is a constant drift along the positive direction. In both (i) and (ii), we provide analysis with and without the presence of an absorbing boundary. 

%%%%%%%%%%%%%%%%%%%%%%%%%%%%%%%%%%%%%%%%%%%%%%%%%
\section{Local Time Statistics: General Formulation}
\l{methods}
\noindent
Let $\{x(\tau);~0\le \tau \le t\}$ represent the stochastic trajectory of the Brownian particle over the time interval $[0,t]$ starting from $x_0$ at time zero. The local time density spent 
by the particle at position $y$ during 
this time interval is given by
\begin{alignat}{1}
L(y,t)=\int_{0}^{t}d\tau ~\delta[x(\tau)-y]~.
\end{alignat}
The appearance of the delta function in the above definition is understood in the following limiting sense
\bea
L(y,t)= \lim_{a \to 0}\frac{T_{2a}(y)}{2a},~~~\text{where}~~T_{2a}(y)=\int_0^t d \tau~[\Theta(x(\tau)-y-a)-\Theta(x(\tau)-y+a)]~,
\eea
where $\Theta(x)$ is the Heaviside step function. In what follows, such regularization procedure should precede every use of the delta function, but we will omit it for brevity. Clearly, $T_{2a}(y)$ measures the occupation time spent by the particle inside the box $[y-a,y+a]$ till time $t$.
In addition, note that by definition $L(y,t)dy$ represents the total time 
spent inside the domain $[y-\frac{dy}{2},y+\frac{dy}{2}]$ up to time 
$t$, and thus is normalized as $\int dy L(y,t)=t$. Given one sample trajectory that started at $x_0$, the local time $L(x_0,t)$ can be computed by summing over the times spent in $x_0$'s vicinity (Fig. \ref{fig:localtime}b).
Here, we would like to compute the probability distribution $P_r(L)$ of $L(y,t)$ where the subscript $r$ represents the presence of resetting. 
Naturally, $P_0(L)$ denotes the probability distribution of $L$ in absence of resetting i.e., when $r=0$.

Since in some of our setups we have an absorbing boundary at the origin,  we would formulate the problem of finding $P_r(L)$ on the ensemble of paths or trajectories which survive till time $t$. In such cases, $P_r(L)$ represents the distribution of $L(y,t)$ conditioned on survival. Hence, 
in what follows, the
survival probability $Q_r(y,t)$ of
a particle which starts from $y$ and is not absorbed till time $t$ will play an important role.
Detailed discussion of the survival probability of diffusion with resetting and how to compute it is given in \aref{survival-MFPT}. 

To compute $P_r(L)$, it will prove convenient to look at the moment generating function of $L(y,t)$. This is given by 
\bea
\mathcal{Q}_r(x,y,k,t)
=\langle e^{-k L(y,t)} \rangle=\int_0^\infty dL~e^{-k L(y,t)}~P_r(L)~, \label{MGF}
\eea
where the average is taken over the trajectories which start at $x$
and $y$ is the position where the local time $L$ is calculated. The moment generating function $\mathcal{Q}_r(x,y,k,t)$ of the local time conditioned on survival can also be written as
\bea
\mathcal{Q}_r(x,y,k,t)=\langle e^{-k L(y,t)} \rangle
=\frac{G_r(x,y,k,t)}{Q_r(x,t)}~,
\label{cond-uncond-GF}
\eea
where the function $G_r(x,y,k,t)$  in the numerator is the moment generating function  of the local time of surviving trajectories. Note, however, that this is not normalized properly. Thus to obtain the moment generating function $\mathcal{Q}_r(x,y,k,t)$ which is conditioned on the survival, one needs to weight $G_r(x,y,k,t)$ with respect to the survival measure $Q_r(x,t)$.  A bit more formally, let us introduce an indicator function $J_t$
which takes the value one if the particle has never visited the origin till time t and zero otherwise.
If $P(L_t,J_t=1)dL_t$ represents the joint probability that local time is between $L_t$ and $L_t+dL_t$ and $J_t=1$. We thus have $G_r(x,y,k,t) = \int dL_t \exp(-k L_t) P_r(L_t,J_{t=1})$. In the $k \to 0$ limit, we get $G_r(x,y,k,t)|_{k \to 0} = Q_r(x,t)$ which ensures normalisation of $\mathcal{Q}_r(x,y,k,t)$ in the $k \to 0$ limit in Eq.~\eqref{cond-uncond-GF}.

We will now derive a backward master equation for $G_r(x,y,k,t)$.
Consider $G_r(x,y,k,t+\Delta t)$ up to time $t+\Delta t$, where $\Delta t$ is a small time increment and split the time interval $[0,t+\Delta t]$ into two parts: $[0,\Delta t]$ and $[\Delta t,t+\Delta t]$. In the first interval 
$\Delta t$, two events can happen: The particle may be reset to $x_0$ with probability $r \Delta t$ or diffuse to a new position $(x+dx)$ with probability $(1-r \Delta t)$ where $dx$ is the stochastic displacement in time $ \Delta t$. Hence, in the subsequent interval $[\Delta t,t+\Delta t]$, the particle starts either from $x_0$ (the former case) or from $x+dx$ (the latter case).
Taking into account all these possibilities, we have 
\bea
G_{r}(x,y,k,t+\Delta t)=r\Delta tG_{r}(x_0,y,k,t)
+(1-r\Delta t)e^{-k \mathcal{U}(x)\Delta t}\langle G_{r}(x+\Delta
x,y,k,t)\rangle_{\Delta x}, ~~~\text{with}~~~\mathcal{U}(x)=\delta(x-y)~,
\label{Gr-dt}
\eea
where the expectation on the second term is taken on different noise realizations \cite{restart_conc17,Hugo-Satya}. We note that the term $\exp(-k\mathcal{U}(x)\Delta t)$ above is somewhat ambiguous since powers of a delta function are not well defined. However, here we will only be interested in $\exp(-k\mathcal{U}(x)\Delta t)$ in the limit $\Delta t \to 0$ and thus approximate it by expanding it to first order in $\Delta t$. Thus, no high powers of the delta function are involved and this gives the same result as 
working with a regularized representation of the delta function as discussed above.
Expanding all other terms in the above equation
to leading order in $\Delta t$, and using $\langle\Delta x\rangle=F(x)\Delta t$ and $\langle(\Delta
x)^{2}\rangle=2D\Delta t$, \eref{Gr-dt} reduces to the following equation in the limit $\Delta t\to0$
\bea
\frac{\partial G_{r}(x,y,k,t)}{\partial
t}=D\frac{\partial^{2}G_{r}(x,y,k,t)}{\partial
x^{2}}+F(x)\frac{\partial G_{r}(x,y,k,t)}{\partial
x}
-[k~ \mathcal{U}(x)+r]G_{r}(x,y,k,t)+rG_{r}(x_0,y,k,t)~,
\l{eq:Qp-eqn}
\eea
with the initial condition $G_{r}(x,y,k,0)=1$.
In order to solve Eq.~\eqref{eq:Qp-eqn}, it is natural to take the Laplace transform 
\bea
\tilde{G}_r(x,y,k,\alpha)&=&\int_0^\infty~dt~e^{-\alpha t}~G_{r}(x,y,k,t)~,
\label{LT-def}
\eea
of both sides of Eq.~\eqref{eq:Qp-eqn} to get
\bea
D\frac{\partial^{2}\tilde{G}_{r}(x,y,k,\alpha)}{\partial
x^{2}}+F(x)\frac{\partial \tilde{G}_{r}(x,y,k,\alpha)}{\partial
x} 
-[\alpha+k~ \mathcal{U}(x)+r]\tilde{G}_{r}(x,y,k,\alpha)+r\tilde{G}_{r}(x_0,y,k,\alpha)~=-1~.
\label{GF-Reset-LT}
\eea
\eref{GF-Reset-LT} is difficult to solve for an arbitrary force $F(x)$.
However, as has been shown in 
\cite{restart_conc17, Hugo-Satya}, the solution $\tilde{G}_{r}(x,y,k,\alpha)$ in the presence of resetting can be written in terms of the solution $\tilde{G}_{0}(x,y,k,\alpha)$ in the absence of resetting as
\bea
\tilde{G}_r(x,y,k,\alpha)=\frac{\tilde{G}_0(x,y,k,\alpha+r)}{1-r\tilde{G}_0(x_0,y,k,\alpha+r)}~,
\label{renewal-formula}
\eea
where the function $\tilde{G}_{0}(x,y,k,\alpha)$ satisfies the backward master equation
\bea
D\frac{\partial^{2}\tilde{G}_{0}(x,y,k,\alpha)}{\partial
x^{2}}+F(x)\frac{\partial \tilde{G}_{0}(x,y,k,\alpha)}{\partial x} 
-[\alpha+k~ \mathcal{U}(x)]\tilde{G}_{0}(x,y,k,\alpha)~=-1~,
\label{GF-noReset-LT}
\eea
with appropriate boundary conditions. For example, in the case of an infinite domain  the boundary conditions are $\tilde{G}_0(x \to \pm \infty, y,k, \alpha) = \frac{1}{\alpha}$.
Indeed, if the initial position $x \to \pm \infty$, the particle will not be able to reach
the location $y$ in finite time, which means that the local time $L(y,t)=0$. Therefore, using the definition
in \eref{cond-uncond-GF} with $r=0$, and setting $Q_0(x\to \pm \infty,t)=1$ since there are no absorbing boundaries, one has $\langle e^{-kL(y,t)} \rangle=1$. The above boundary condition then follows immediately. On the other hand, if there is an absorbing boundary at the origin, we use the following boundary conditions: $\tilde{G}_0(x \to + \infty, y,k, \alpha) = \frac{1}{\alpha}$ and 
 $\tilde{G}_0(0,y,k,\alpha)=0$. The former condition follows from the same argument as mentioned above. However the latter boundary condition has to be understood more carefully. To see this first note that $G_0(x,y,k,t)$ (in \eref{cond-uncond-GF} with $r=0$) can be written formally in terms of the Brownian path integral
 \bea
G_0(x,y,k,t) \propto \int_0^\infty d\tilde{x} \int_{x(0)=x}^{x(t)=\tilde{x}}~\mathcal{D}[x(\tau)]\exp \left(-\int_0^t d\tau \left[\frac{1}{4D} \left(\frac{dx}{d\tau}+F(x) \right)^2+ k~\mathcal{U}(x(\tau))\right] \right) \prod\limits_{\tau=0}^{t} \theta[x(\tau)]~,
\eea
where $\theta[x(\tau)]=1$ for $x(\tau)>0$ and is zero otherwise.
It is therefore clear from the definition above that $G_0(x\to 0,y,k,t)=0$, since the probability weight of those trajectories which start at the absorbing boundary $x=0$ and survive till time $t$ identically vanishes. Hence, the path integral also vanishes. Consequently, its Laplace transform  $\tilde{G}_0(0,y,k,\alpha)=0$, which is our second boundary condition in the presence of an absorbing boundary at the origin.

 \eref{renewal-formula} offers an interesting renewal interpretation \cite{restart_conc17, Hugo-Satya}, which becomes evident once 
the right hand side is expanded in series as $\tilde{G}_r(x,y,k,\alpha)=\tilde{G}_0(x,y,k,\alpha+r) \sum_{n=0} r^n \tilde{G}_0^{n}(x_0,y,k,\alpha+r)$. The $n$-th term contains contributions from those trajectories which have been reset $n$ times during the time interval $[0,t]$ and the time convolution of the $G_0$ propagators in between resetting events becomes a product in the Laplace domain.

After one gets $\tilde{G}_r(x,y,k,\alpha)$, one can in principle obtain the moment generating function $G_{r}(x,y,k,t)$ by performing the inverse Laplace transform. However, this task is usually hard for finite time $t$ and arbitrary $F(x)$. Instead we look at the moments of the local time $\langle L^n(y,t)  \rangle$ which turn out to be easier to compute. These moments are related  directly to the derivatives of the generating function (with respect to $k$) as 
\bea
\langle L^n(y,t)  \rangle = \frac{1}{Q_r(x,t)}\frac{\partial^n}{\partial (-k)^n} G_r(x,y,k,t)\bigg|_{k\to 0} \equiv \frac{\langle l^n(y,t)  \rangle}{Q_r(x,t)}~,
\label{lt}
\eea
where we have defined the auxiliary moments as $\langle l^n(y,t)  \rangle \equiv  \frac{\partial^n}{\partial (-k)^n} G_r(x,y, k,t)|_{k\to 0}$. Now using Eq.~\eqref{LT-def}, it is easy to see that $\langle l^n(y,t) \rangle$ is given by the inverse Laplace transform of the $n$-th derivative of  $\tilde{G}_r(x,y,k,\alpha)$ with respect to $k$ as follows 
\bea
 \langle l^n(y,t) \rangle&=&\text{Inverse~Laplace~transform}\left[\frac{\partial^n}{\partial (-k)^n} \tilde{G}_r(x,y,k,\alpha)\bigg|_{k\to 0}\right]  \nonumber \\
 &=& \frac{1}{2 \pi i}\int_{\mathcal{B}}d\alpha~\text{e}^{\alpha t}~
 \frac{\partial^n}{\partial (-k)^n} \tilde{G}_r(x,y,k,\alpha)\bigg|_{k\to 0}~,
\label{meanLTfull}
\eea
where we have used Cauchy's inversion formula. Here, $\mathcal{B}$ represents the Bromwich contour on which the integral has to be evaluated.

Throughout the rest of the paper we take the three following points in space: $y$ (where the local time is calculated), the resetting position $x_0$ and the initial position $x$ to be the same point which we henceforth denote by $x_0$. In the following, we present detailed calculations for the distribution and various moments of $L(x_0,t)$ in the absence, and presence, of resetting and an absorbing boundary.

\section{Local time of simple diffusion}
\l{simple}
\subsection{Infinite domain}
\l{infinite}
\noindent
We first consider the simple case of diffusion in an infinite domain in 1D, i.e., a line, without any external force acting on the particle.
As mentioned in the previous section, to get $\tilde{G}_r(x,x_0,k,\alpha)$ in presence of resetting, one first needs to solve the problem without resetting i.e., Eq.~\eqref{GF-noReset-LT} with boundary conditions $\tilde{G}_0(x \to \pm \infty,x_0,k,\alpha)=\frac{1}{\alpha}$ and then finally set $x=x_0$. Note that in this case $Q_r(x_0,t)=1$. The problem in the absence of resetting was considered in \cite{Sabhapandit:2006}, where the authors found that the generating function at $x_0=0$ is given by $\tilde{G}_0(0,0,k,\alpha)=\frac{1}{\sqrt{\alpha}(\sqrt{\alpha}+k/\sqrt{4D})}$. However, it can be shown that for any non-zero $x_0$ one still finds $\tilde{G}_0(x_0,x_0,k,\alpha)=\frac{1}{\sqrt{\alpha}(\sqrt{\alpha}+k/\sqrt{4D})}$, which is independent of $x_0$. This happens because of the translational invariance of the problem.
Substituting $\tilde{G}_0(x_0,x_0,k,\alpha)$  in Eq.~\eqref{renewal-formula}, we find 
\bea
\tilde{G}_r(x_0,x_0,k,\alpha)=\frac{1}{\alpha+k\sqrt{\alpha+r}/\sqrt{4D}}~,
\label{GF-1}
\eea
which is again independent of $x_0$ because the resetting position was chosen to be identical to the initial position. 
Using this expression in Eqs.~\eqref{lt} and \eqref{meanLTfull}, it is easy to see that the Laplace transform of the $n$-th moment $\langle L^n(x_0,t) \rangle$ is given by
\bea
\int_0^\infty~dt~e^{-\alpha t} \langle L^n(x_0,t) \rangle=\frac{\partial^n}{\partial (-k)^n} \tilde{G}_r(x_0,x_0,k,\alpha)\bigg|_{k\to 0}=\frac{n!}{(4D)^{n/2}} \frac{(\alpha+r)^{n/2}}{\alpha^{n+1}}~,~~\text{for}~~n=1,2,3,...
\eea
which can be inverted to provide $\langle L^n(x_0,t) \rangle$ for $n \ge 1$ and we find
\bea
\langle L^n(x_0,t) \rangle &=& \frac{e^{-r t}}{(4D)^{n/2}} \left[\frac{\partial^n}{\partial z^n}z^{n/2}e^{zt} \right]_{z=r}  \nonumber \\
&-&  
\begin{cases}
0 & \text{for}~n=2m \\
\frac{n!}{(4D)^{n/2}} \frac{(-1)^mn \Gamma \left({n}/{2}\right)}{2\pi r^{n/2}}  e^{-r t}~U\left(\frac{n}{2}+1,1-\frac{n}{2},r t\right) &\text{for}~n=2m+1~,
\end{cases}
~~m=1,2,3,...
\eea
where $\Gamma(x)$ is the Gamma function and $U(a,b,x)$ is the Tricomi hypergeometric function \cite{Ryzhik,Stegun}.
The first three moments are given explicitly as follows:
\bea
\langle L(x_0,t) \rangle&=&\sqrt{\frac{t}{4\pi D}}e^{-rt}+\frac{1+2rt}{4\sqrt{rD}} ~\erf[\sqrt{rt}]~, \label{mean-Infinite-r0p6} \\
\langle L^2(x_0,t) \rangle&=&\frac{1}{2D}\left(t+\frac{r t^2}{2}\right)~, \label{2ndM-Infinite-r0p6} \\
\langle L^3(x_0,t) \rangle&=&\frac{ 6\left(\sqrt{\pi } \left(8 r^3 t^3+36 r^2 t^2+18 r t-3\right)
   \text{erf}\left(\sqrt{r t}\right)+2 e^{-r t} \sqrt{r t} \left(4 r^2 t^2+16 rt+3\right) \right) }{48 \sqrt{\pi } (4Dr)^{3/2}}~. 
 \label{3rdM-Infinite-r0p6}
\eea
We also compute the third cumulant given by $\mathcal{C}_3=\langle L^3(x_0,t) \rangle-3\langle L(x_0,t) \rangle\langle L^2(x_0,t) \rangle+2\langle L(x_0,t) \rangle^3$
and this gives $\mathcal{C}_3 \sim \frac{t}{\sqrt{rD^3}}$, which scales linearly with $t$ at a large time as expected.
It is easy to see that in the $r \to 0$ limit the above expressions correctly reproduce the results for free diffusion in the absence of resetting as reported in \cite{Sabhapandit:2006}.
In addition, we note that in the long time limit the mean local time scales as
\bea
\langle L(x_0,t) \rangle \sim \frac{1}{2}\sqrt{\frac{r}{D}}t~.
\eea
Thus, the empirical density obeys $\lim_{t \to \infty} \frac{\langle L(x_0,t) \rangle}{t} = \frac{1}{2}\sqrt{\frac{r}{D}}$,
which is equal to the probability density of finding the particle at $x=x_0$ in the steady state in the presence of resetting \cite{Restart1}.

The full distribution of $L(x_0,t)$ can be also computed exactly. In order to do that, 
we first perform the inverse Laplace transform of \eref{GF-1} with respect to $k$ to find 
\bea
\tilde{F}_r(x_0,x_0,L, \alpha)
=\frac{\sqrt{4D}}{\sqrt{\alpha+r}}~\exp \left[-\sqrt{4D} \frac{\alpha L}{\sqrt{\alpha+r}} \right]~.
\label{aux-1}
\eea
Next, we perform the Laplace inversion with respect to $\alpha$. However, since this calculation is lengthy and cumbersome, we simply state the result here while leaving the details to \aref{LTInfinfite}.
Using the fact that $Q_r(x_0,t)=1$ in the absence of an absorbing boundary, we find
\bea
P_r(L)&=&e^{-rt}\left[ \sqrt{\frac{4D}{\pi t}}e^{-DL^2/t} + \sqrt{4D} ~\sum_{j=1}^{\infty} \frac{(\sqrt{4D}~ rL)^{j}}{j! \Gamma(\frac{j}{2})} H_j\left(\frac{DL^2}{t},t \right)\right]~,
\label{PDF-Infinite-r0p6} 
\eea
where we have defined the functions $H_j(m,t)$ (see \aref{LTInfinfite})
\bea
H_j(m,t)&=&t^{\frac{j}{2}-1} \Bigg( \frac{\Gamma(\frac{j}{2})~{}_{1}{\cal F}_{1}[\frac{1-j}{2},\frac{1}{2},-m]}{\Gamma(\frac{1+j}{2})} 
-2\sqrt{m}~ {}_{1}{\cal F}_{1}[\frac{2-j}{2},\frac{3}{2},-m]  \Bigg)~,
\eea
and ${}_{1}{\cal F}_{1}[a,b,x]$ is the confluent hyper-geometric function which is different than the Tricomi hypergeometric function defined earlier \cite{Ryzhik,Stegun}. Note that this result is similar to a result for the positive occupation time $\int_0^\infty dy L(y,t)$ which was studied in great detail in \cite{Hugo-Satya}.
The exact result in \eref{PDF-Infinite-r0p6} is in excellent agreement with the simulations as shown in \fref{fig:lt-inft-pdf-mean-rp6}a. 
For $r=0$, we observe that only the first term in \eref{PDF-Infinite-r0p6} survives which correctly reproduces the result $P_0(L)=\sqrt{\frac{4D}{\pi t}}e^{-DL^2/t}$ reported earlier 
in \cite{Sabhapandit:2006}. 

For $r>0$, we observe that in the long time limit one expects a large number of resetting events, typically of the order of $\sim rt$, to occur within the observation time $t$. On the other hand, the motion of the particle after a resetting event is independent of its motion prior to the resetting event. As a result, the local times spent at $x_0$ in between two successive resetting events are also independent. This means that the total local time spent at $x_0$ is basically a sum of the local times spent in between resetting events i.e., $L(x_0,t) \approx L_1+L_2+...+L_{\lfloor rt \rfloor}$. Since in our case the resetting position and the position where we compute the local time are the same, the $L_i$-s are also identically distributed. Hence due to the central limit theorem,  the distribution of $L(x_0,t)$ is asymptotically Gaussian 
\bea
P_r(L) \simeq \frac{1}{\sqrt{2 \pi \sigma^2_r(t)}}\text{exp}\left(-\frac{(L-\langle L(x_0,t) \rangle)^2}{2 \sigma^2_r(t)} \right)~,
\label{Gaussian-infinite}
\eea 
where the mean $\langle L(x_0,t) \rangle $ and the variance $\sigma^2_r(t)\equiv \langle L^2(x_0,t) \rangle-\langle L(x_0,t) \rangle^2$ can be computed from the long time limit of Eqs.~\eqref{mean-Infinite-r0p6} and \eqref{2ndM-Infinite-r0p6} and are given explicitly by $\langle L(x_0,t) \rangle=\frac{1}{2}\sqrt{\frac{r}{D}}t$ and $\sigma_r^2(t)=\frac{t}{4D}-\frac{1}{16r D}$. This approximate form is verified numerically in \fref{fig:lt-inft-pdf-mean-rp6}b. Notably, a similar Gaussian form for the distribution of the local time was also observed in the absence of resetting for Brownian motion confined in a domain with reflecting boundaries \cite{RBM}.

\begin{figure}[t]
\includegraphics[width=6.8cm]{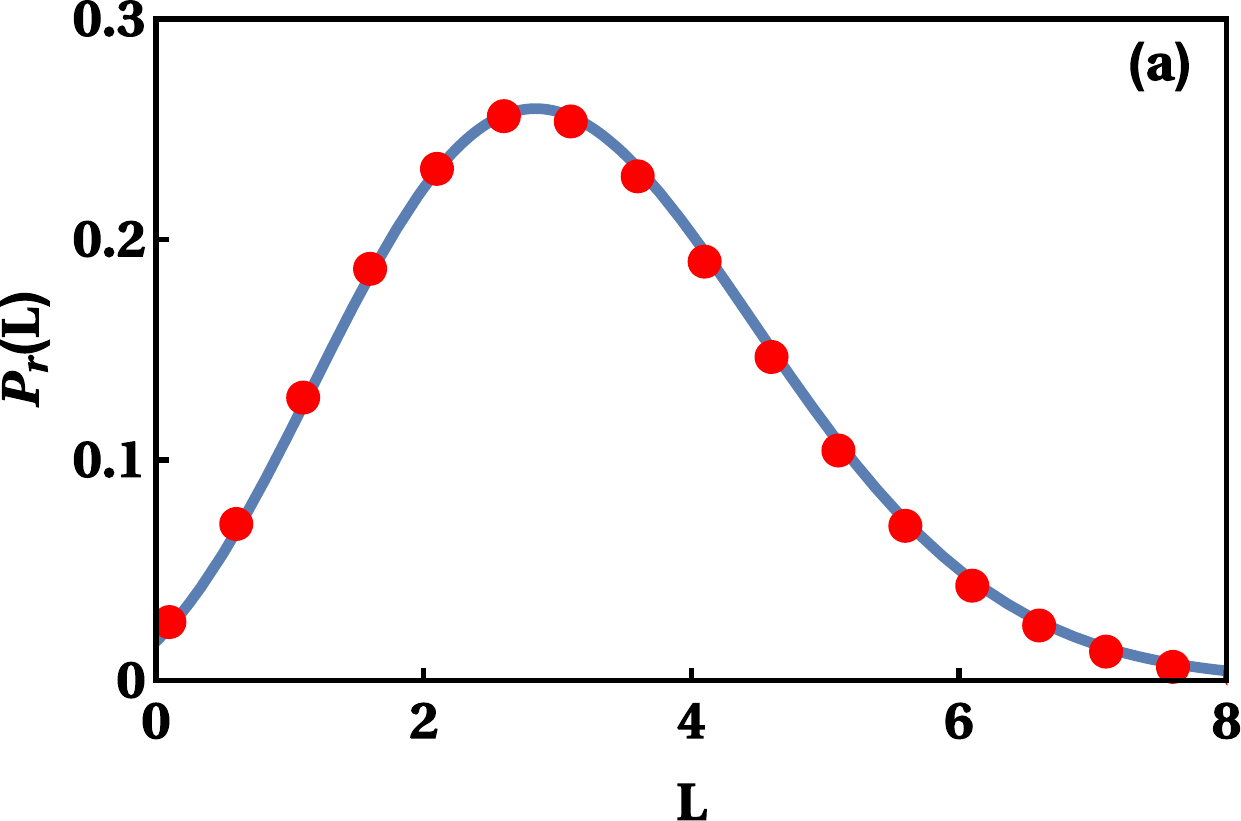}
\includegraphics[width=7cm]{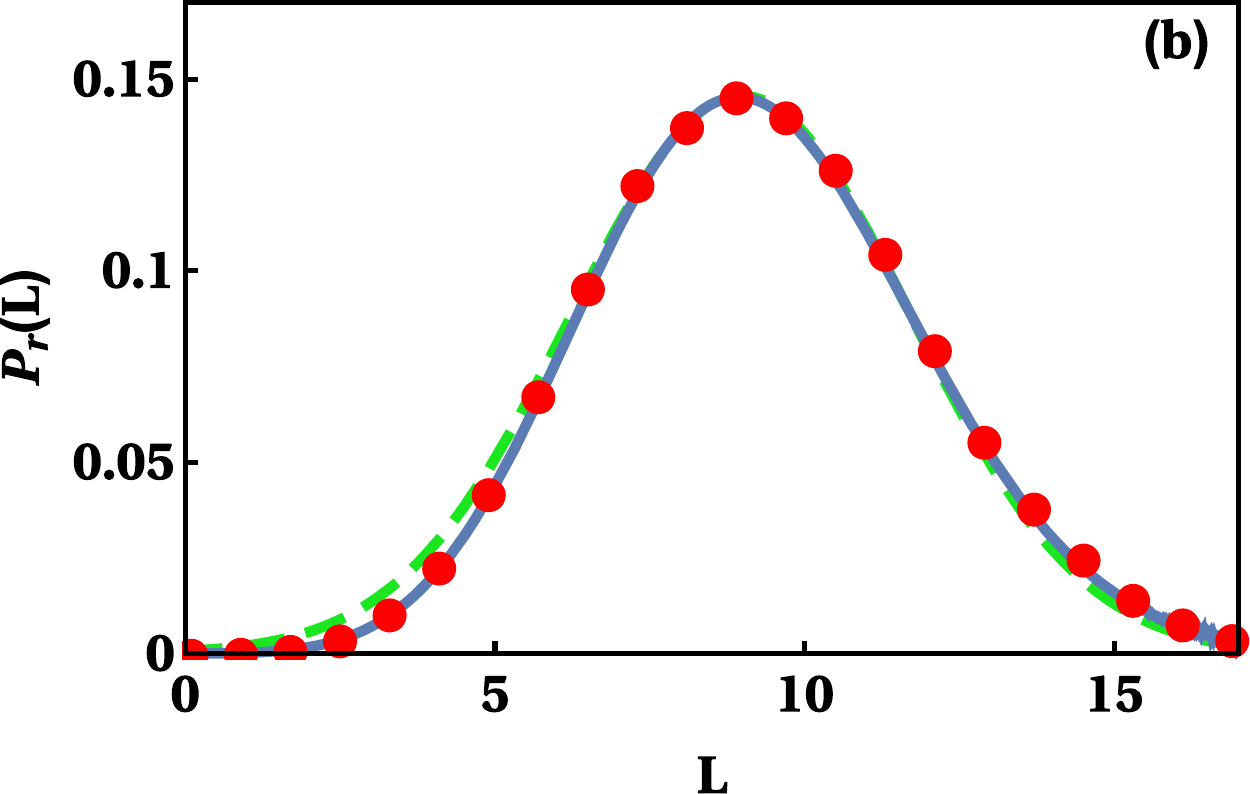}
\caption{The probability distribution $P_r(L)$ of the local time for simple diffusion on a line in the absence of drift: (a) Solid blue line is the exact formula from \eref{PDF-Infinite-r0p6} and
full red circles come from a simulation. Here,  $r=0.6,D=0.5$ and $t=5$. (b) The Gaussian approximation from \eref{Gaussian-infinite}  (dashed green line) is in very good agreement with 
simulations (full red circles) and the exact solution from \eref{PDF-Infinite-r0p6} (solid blue line). Here, $r=0.7,D=0.5$ and $t=15$.}
\l{fig:lt-inft-pdf-mean-rp6}
\end{figure}

\subsection{Semi-infinite domain with an absorbing boundary at the origin}
\l{semi-infinite-no-drift}
\noindent
We now consider the case of an absorbing boundary at the origin. As before, the particle resets to its initial position $x_0$ which is also the place where we compute the local time.
For a simpler presentation, we examine the cases with and without resetting separately.

\subsubsection{No Resetting: $r=0$}
\l{semi-infinite-no-drift-req0}
\noindent
We start from \eref{GF-noReset-LT} with $F(x)=0$. Since we have an absorbing boundary at $x=0$, we solve this equation  with the following boundary conditions
\bea
\tilde{G}_0(x \to \infty,x_0,k,\alpha) = \frac{1}{\alpha}~,~~~~~\tilde{G}_0(0,x_0,k,\alpha)=0~.
\eea
The solution is given by
\bea
\tilde{G}_0(x_0,x_0,k,\alpha)=\frac{1}{\alpha} \left[ 1-\frac{\beta D+k \sinh(\beta x_0)}{\beta D e^{\beta x_0}+k\sinh(\beta x_0)}  \right]~,
\label{GF-2}
\eea
where $\beta=\sqrt{\frac{\alpha}{D}}$.
Performing the inverse Laplace transform with respect to $k$ first, we get 
\bea
\tilde{F}_0(x_0,x_0,L,\alpha)=\frac{\mathrm{cosech}(\beta x_0)}{\beta}\left( -1+e^{\beta x_0} \right)~\exp \left[-\beta D L ~ \mathrm{cosech}(\beta x_0) e^{\beta x_0} \right]~.
\label{aux-SI}
\eea
We still have to perform the inverse Laplace transform (ILT) with respect to $\alpha$. However, this turns out to be difficult for finite $t$. We instead examine the ILT in different limits. In the small 
$t$ limit which corresponds to a large $\alpha$ expansion, we find 
$\tilde{F}_0(x_0,x_0,L,\alpha) \sim \sqrt{\frac{4D}{\alpha}} e^{-\sqrt{4D\alpha}L}$. Inverting this we obtain the density of the local time at short times 
$P_0(L)=\sqrt{\frac{4D}{\pi t}}e^{-DL^2/t}$ \cite{Sabhapandit:2006}. It is important to note that this expression is identical to the expression for the local
time density in the infinite domain \cite{Sabhapandit:2006}, because at short times the system is yet to experience the boundary at  $x=0$. On the other hand, the
large $t$ limit corresponds to small values of $\alpha$ for which we
find $\tilde{F}_0(x_0,x_0,L,\alpha)\sim \sqrt{\frac{D}{\alpha}}e^{-DL/x_0}$. 
Laplace inversion of this expression with respect to $\alpha$ gives $\sqrt{\frac{D}{\pi t}}e^{-DL/x_0}$. 

\begin{figure}[t]
\includegraphics[width=.465\hsize]{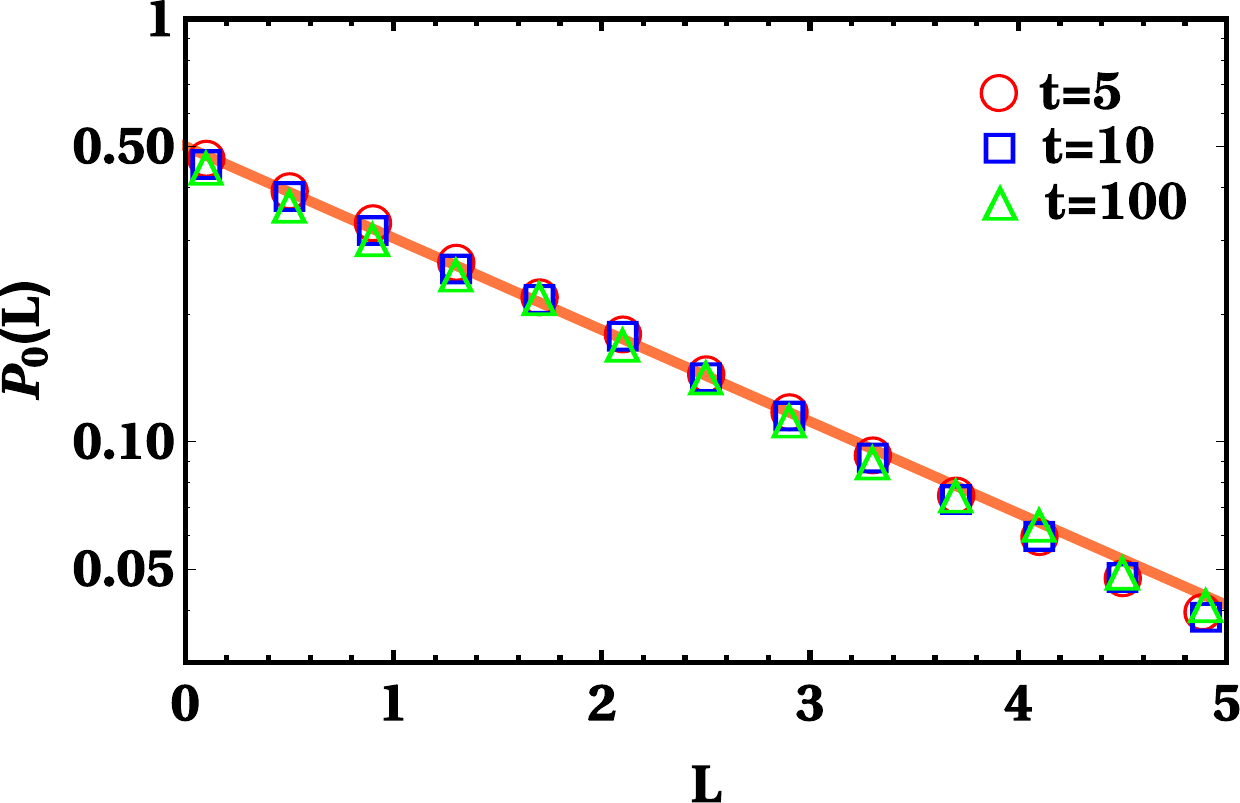}
\caption{The long time asymptotics of the probability distribution of the local time from \eref{PDF-semi-Infinite-r0} (solid orange line) and results from numerical simulations (symbols). Here, $x_0=1$ and $D=0.5$. Data collapse of the simulation results for $t=5$ (red circles), $t=10$ (blue squares) and $t=100$ (green triangles) verify that in the absence of resetting the probability distribution of the local time is asymptotically independent of the observation time.}
\label{lt-semiinft-pdf-l0-r0}
\end{figure}

We now recall that the density of the local time is measured conditioned on survival till time $t$. The survival probability for the case analyzed herein is a classical result from probability theory and is given by
$Q_0(x_0,t)=\erf(\frac{x_0}{\sqrt{4Dt}})$ \cite{Satya-current-science,Schehr-review,RednerBook,Hughes-book}. In the long time limit, this survival probability decays as $Q_0(x_0,t)|_{t \to \infty} \approx \frac{x_0}{\sqrt{\pi D t}}$.  
Hence, the distribution of the local time density in this limit is given by
\bea
P_0(L) = \frac{1}{Q_0(x_0,t)} \left[ \sqrt{\frac{D}{\pi t}}e^{-DL/x_0} \right]
=\frac{D}{x_0}e^{-\frac{D L}{x_0}}~,
\label{PDF-semi-Infinite-r0}
\eea
which turns out to be independent of the observation time. The result in \eref{PDF-semi-Infinite-r0} is numerically verified in \fref{lt-semiinft-pdf-l0-r0}.

%%%%%%%%%%%%%%%%%%%%%%%%%%%%%%%%%%%%%%%%%%%%%%%%%%%%%%%%%%
%%%%%%%%%%%%%%%%%%%%%%%%%%%%%%%%%%%%%%%%%%%%%%%%%%%%%%%%%%%

\subsubsection{With resetting: $r > 0$}
\l{semi-infinite-no-drift-rneq0}
\noindent
We now consider the $r > 0$ case, where the moment generating function $\tilde{G}_r(x_0,x_0,k,\alpha)$ of the local time is obtained by substituting $\tilde{G}_0(x_0,x_0,k,\alpha)$ from \eref{GF-2} into \eref{renewal-formula} to get
\bea
\tilde{G}_r(x_0,x_0,k,\alpha)=\frac{1-\exp \left[ -\sqrt{\frac{r+\alpha}{D}}~x_0 \right]}{\alpha+r \exp \left[ -\sqrt{\frac{r+\alpha}{D}}~x_0 \right]+\frac{k}{2} \sqrt{\frac{r+\alpha}{D}}\left( 1-\exp\left[-2 \sqrt{\frac{r+\alpha}{D}}~x_0 \right] \right)}~.
\label{MGF-no-drift-w-reset}
\eea
In addition, we note that in this case the survival probability in Laplace space is given by \cite{Restart1} (also see \aref{survival-MFPT} for a derivation of this formula)
\bea
\tilde{Q}_r(x_0,\alpha)=\frac{1-\exp \left[- \sqrt{\frac{r+\alpha}{D}}~x_0 \right]}{\alpha+r \exp \left[- \sqrt{\frac{r+\alpha}{D}}~x_0 \right]}~,
\label{qrSI}
\eea
which can be inverted by the Bromwich integral $Q_r(x_0,t)= (1/2 \pi i)~\int_{\mathcal B}d\alpha~e^{\alpha t} \tilde{Q}_r(x_0,\alpha)$. 
At short times this integral gets its dominant contributions from large $\alpha$-s, where one can approximate
\bea
\tilde{Q}_r(x_0,\alpha) \sim \frac{1}{\alpha}\left( 1-\exp \left[- \sqrt{\frac{\alpha}{D}}~x_0 \right]  \right)~.
\eea
Performing Laplace inversion of this expression 
immediately provides the survival probability
\bea
Q_r(x_0,t) \sim \erf \left(\frac{x_0}{\sqrt{4Dt}}\right)~,
\eea
in the absence of resetting, as expected.  
To evaluate $\tilde{Q}_r(x_0,\alpha)$ for arbitrary $t$, one has to consider an appropriate contour integral and then use Cauchy's residue theorem. 
The appropriate contour in our case is drawn in Fig.~\ref{contour_mt}. 
Here we will only state the result while leaving the proof in \aref{ILT-derivation}. After performing the contour integral, we obtain
\bea
Q_r(x_0,t) &=& \frac{(1-e^{-b\sqrt{u_0}})}{1-(b/2\sqrt{u_0})~e^{-b\sqrt{u_0}}}e^{-r u_0 t}+h_r(b,t)~,\label{surv-no-drift-w-reset}\\\text{with~~~}
h_r(b,t) &=&\frac{2e^{-rt}}{\pi}~\int_0^\infty dv~e^{-rtv^2}~\frac{v^3\sin(bv)}{v^4+2+2(1+v^2)[1-\cos(b v)]}~,
\eea
where $u_0 (0 \le u_0 \leq 1)$ is a solution of the equation $u_0=1-\exp \left( -\sqrt{u_0} b\right)$ and $b=\sqrt{\frac{r}{D}}x_0$.
Evidently in the large $t$ limit, the first term in Eq.~\eqref{surv-no-drift-w-reset} dominates and the survival probability decays as $Q_r(x_0,t) \simeq e^{-r u_0 t}$. \\

\begin{figure}[t]
\includegraphics[width=5cm,height=3.5cm]{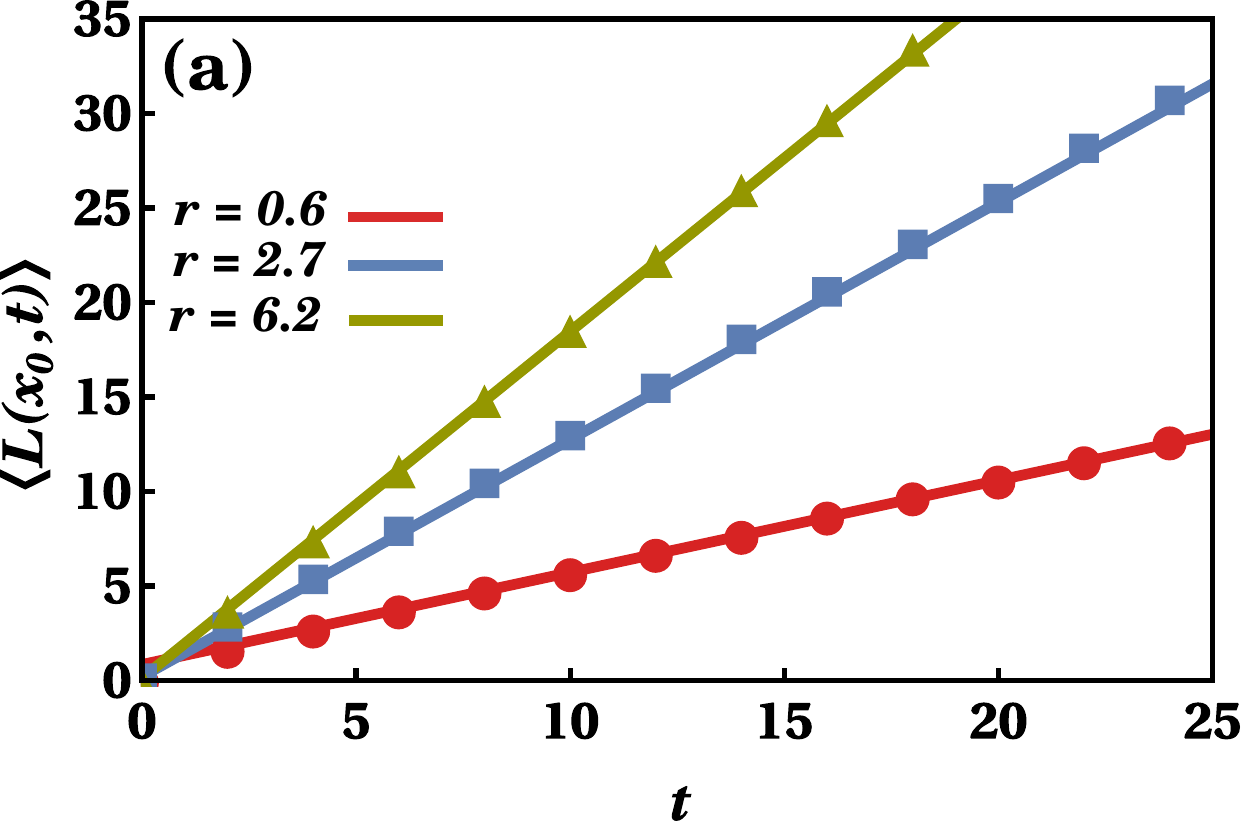}
\includegraphics[width=5cm,height=3.5cm]{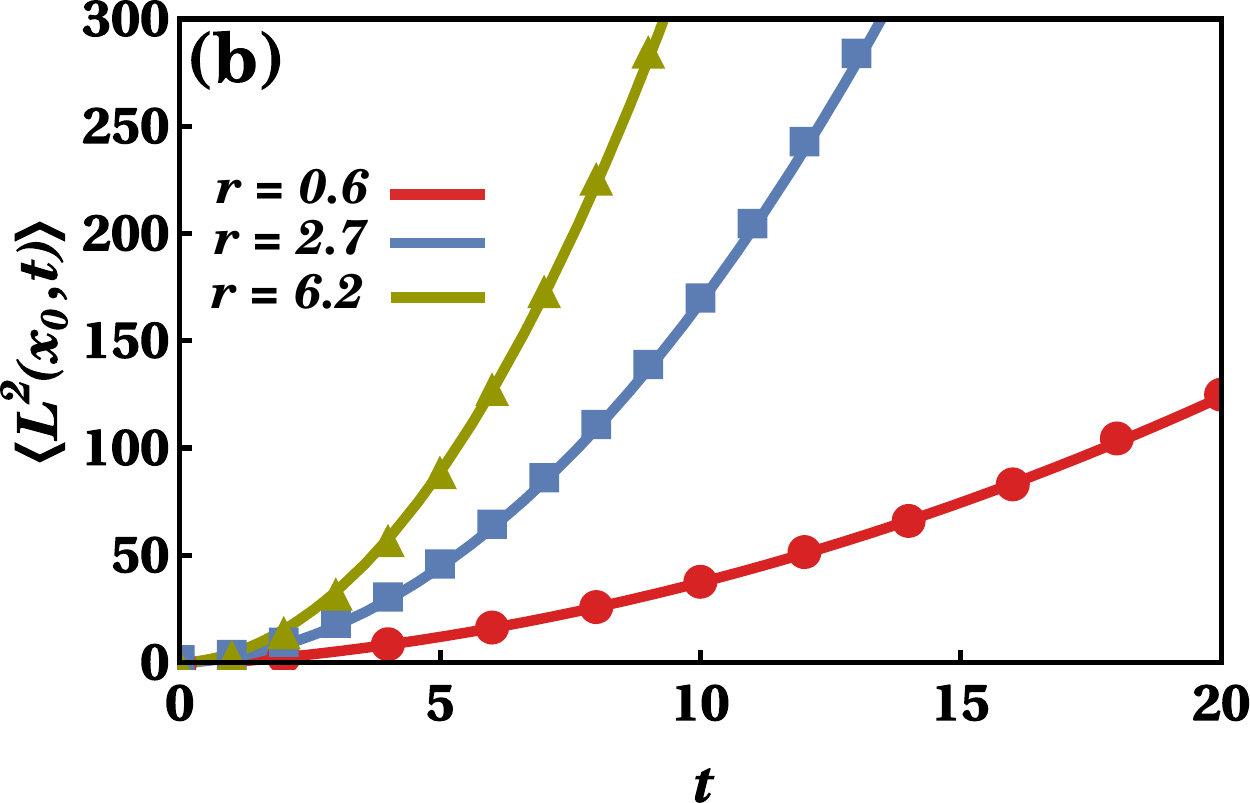}
\includegraphics[width=5cm,height=3.45cm]{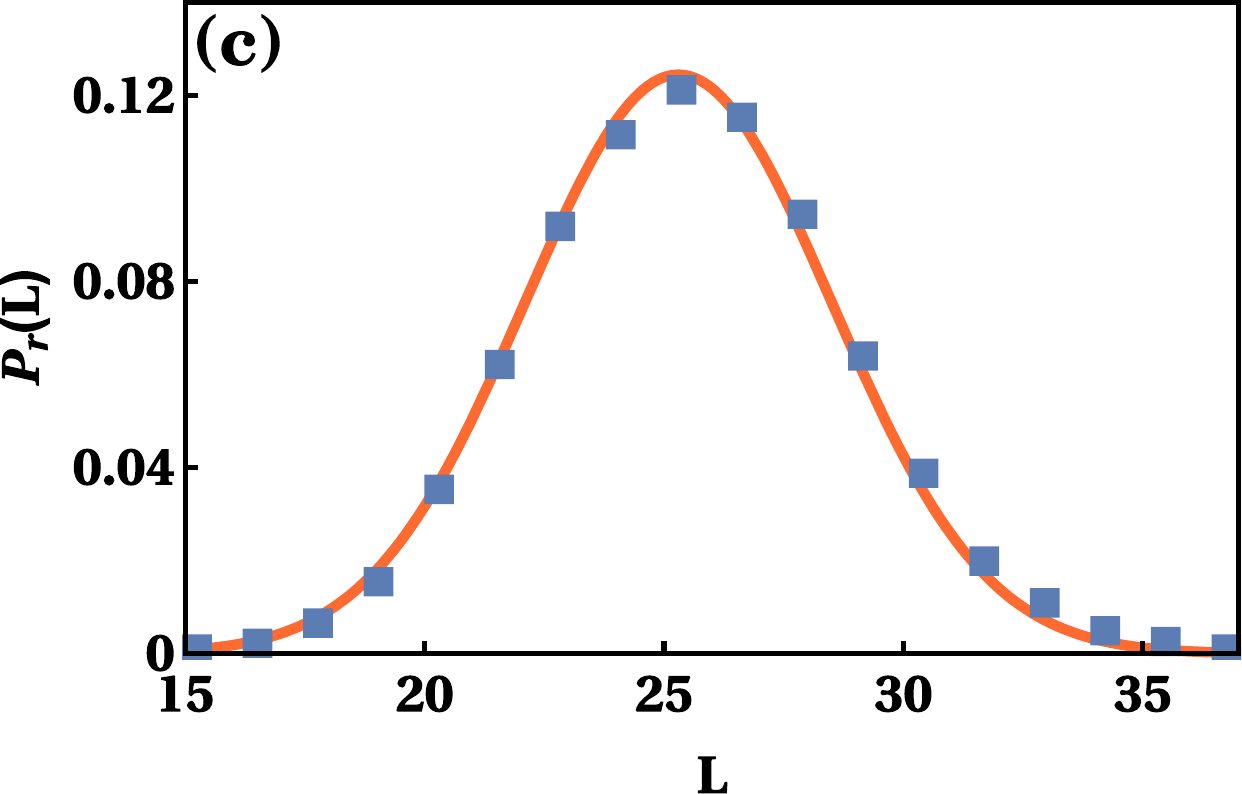}
\caption{The mean (a) and the second moment (b) of the local time of a Brownian particle in a semi-infinite domain with an absorbing boundary at the origin. Solid lines come from Eqs. (\ref{meanlt-conditioned}) and (\ref{smlt-conditioned}) and symbols come from numerical simulations. Here, $x_0=1$, $D=0.5$, and plots are made for three different resetting rates: $r=0.6$ (red), $2.7$ (blue) and $6.2$ (green). (c) The Gaussian approximation from \eref{Gaussian-semi-infinite} (solid orange line) and results from a numerical simulation (full squares). Here, $x_0=1,D=0.5,r=2.7$ and $t=20$.}
\label{lt-semiinft-mean-l0}
\end{figure}

%%%%%%%%%%%%%%%%%%%%%%%%%%%%%%%%%%%%%%%%%%%%%%%%%%%%%%%%%%
Let us now turn our attention back to $\tilde{G}_r(x_0,x_0,k,\alpha)$ in Eq.~\eqref{MGF-no-drift-w-reset}. As in the previous case of unbounded diffusion on a line,
inverting the Laplace transform of $\tilde{G}_r$ is hard, whereas, computing the moments turns out to be easier. Starting from Eq.~\eqref{lt} and using Eq.~\eqref{meanLTfull}, we obtain
\bea
\langle L^n(x_0,t)  \rangle &=& \frac{\langle l^n(x_0,t)  \rangle}{Q_r(x_0,t)},~~\text{with}~~
\langle l^n(x_0,t) \rangle=\frac{n!}{2 \pi i}\int_{\mathcal{B}}d\alpha~\text{e}^{\alpha t}~
 \frac{\mathcal{A}_r(n,\alpha)}{[g_r(\alpha)]^{n+1}},~~~\text{where} \\
 \mathcal{A}_r(n,\alpha)&=&\frac{1}{2^{n}}\left( \frac{\alpha+r}{D}\right)^n~\left( 1-\exp\left[-2 \sqrt{\frac{r+\alpha}{D}}~x_0 \right] \right)^n~\left(1-\exp\left[- \sqrt{\frac{r+\alpha}{D}}~x_0 \right] \right)~,~~~\text{and}~ \\
 && ~~~g_r(\alpha)=\alpha+r \exp \left[- \sqrt{\frac{r+\alpha}{D}}~x_0 \right].
\eea
As observed earlier, the function $g_r(\alpha)$ has a zero at $\alpha_0=-r+ru_0$ where $u_0$ satisfies  $u_0=1-\exp \left( -\sqrt{u_0} b\right)$ with $b=\sqrt{\frac{r}{D}}x_0$.
Hence the integrand has an $(n+1)$-th order pole at $\alpha_0$ in addition to a branch point at $\alpha=-r$ (see \fref{contour_mt}). 
Once again the Bromwich integral can be performed by considering the same contour in \fref{contour_mt} which was used to invert \eref{qrSI}.
This provides two contributions: one from the pole (of multiple order) and the other one from the branch cut $(-\infty, -r)$. Hence, one 
can write the following relation for $\langle L^n(x_0,t)  \rangle$
\bea
\langle L^n(x_0,t)  \rangle &=& \frac{\text{Residue~at~}\alpha_0=-r(1-u_0)~~+~~\text{Contribution~from~the~branch~cut~}(-\infty,-r)}{Q_r(x,t)}~.
\eea
which provides exact theoretical estimates for the moments. Though the explicit expressions of these terms can quickly become cumbersome, they can be computed very easily with the aid of a computer. It is clear that the contribution from the residue term decays as $e^{-r u_0 t}$ where the contribution from the integral along branch cut decays faster than $e^{-rt}$. Hence, in the long time limit the residue term dominates. From the same reason, the residue term also provides the leading large $t$ decay of the survival probability $Q_r(x_0,t)$. 
Hence, in the large $t$ limit, we have the following expressions for the first and second moments and for the variance
\bea
\langle  L(x_0,t) \rangle & \simeq &~ \frac{ f_1'(\alpha_0)+t f_1(\alpha_0) }{ \mathcal{I}(b,u_0)} , 
\l{meanlt-conditioned} \\
\langle  L^2(x_0,t) \rangle & \simeq&~ \frac{ f_2''(\alpha_0)+2t f_2'(\alpha_0)+t^2 f_2(\alpha_0)  }{\mathcal{I}(b,u_0)}~,
\l{smlt-conditioned}\\
\sigma^2_r(t)& \simeq &~\frac{  \mathcal{I}(b,u_0) f_2''(\alpha_0)-f_1'(\alpha_0)^2   +2t \left[ \mathcal{I}(b,u_0) f_2'(\alpha_0)-f_1'(\alpha_0)f_1(\alpha_0) \right]  }{\mathcal{I}(b,u_0)^2}~,
\l{varlt-conditioned}
\eea
where
\bea
\mathcal{I}(b,u_0)=\frac{(1-e^{-b\sqrt{u_0}})}{1-(b/2\sqrt{u_0})~e^{-b\sqrt{u_0}}} ~,
\eea
and the functions $f_1(\alpha)$ and $f_2(\alpha)$ can be computed explicitly. These are given by
\bea
f_1(\alpha)&=&\frac{1}{2}\sqrt{\frac{\alpha +r}{D}}~\frac{\left( 1+e^{-\sqrt{\frac{\alpha +r}{D}} x_0} \right)\left( 1-e^{-\sqrt{\frac{\alpha +r}{D}} x_0} \right)^2}{ \left[ 1- (b/2\sqrt{u_0})~e^{-b\sqrt{u_0}}+\frac{\alpha+r-r u_0}{2}~e^{-b\sqrt{u_0}}~\frac{x_0^2}{4Du_0}\left( 1+\frac{1}{b\sqrt{u_0}} \right) \right]^2}~, \label{f1}\\
f_2(\alpha)&=&\frac{1}{4}\frac{\alpha +r}{D}~\frac{\left( 1+e^{-\sqrt{\frac{\alpha +r}{D}} x_0} \right)^2\left( 1-e^{-\sqrt{\frac{\alpha +r}{D}} x_0} \right)^3}{ \left[ 1- (b/2\sqrt{u_0})~e^{-b\sqrt{u_0}}+\frac{\alpha+r-r u_0}{2}~e^{-b\sqrt{u_0}}~\frac{x_0^2}{4Du_0}\left( 1+\frac{1}{b\sqrt{u_0}} \right) \right]^3} ~,
\label{f2}
\eea
where $u_0$
is a solution of the transcendental equation $u_0=1-\exp \left( -\sqrt{u_0} b\right)$, with $b=\sqrt{\frac{r}{D}}x_0$.
The results in \eref{meanlt-conditioned} and \eref{smlt-conditioned} are numerically verified in panels (a) and (b) of \fref{lt-semiinft-mean-l0}. 
Exact expressions for higher moments can be obtained in a similar fashion.

For large $t$, the distribution of $L(x_0,t)$ should be approximately Gaussian. The argument is similar to that brought in the previous section. If a particle was able to survive for a long time $t$, it would typically experience a large number ($\sim rt$) of reset events. Hence, the total local time spent at $x_0$ is a sum of many independent and identically distributed local times that were spent at $x_0$ in between resetting events. The central limit theorem then gives
\bea
P_r(L) \simeq \frac{1}{\sqrt{2 \pi \sigma^2_r(t)}}\text{exp}\left(-\frac{(L-\langle L(x_0,t)  \rangle)^2}{2 \sigma^2_r(t)} \right)~,
\label{Gaussian-semi-infinite}
\eea
where the mean $\langle L(x_0,t)  \rangle$ and variance $\sigma^2_r(t)$ are given by Eqs.~\eqref{meanlt-conditioned} and \eqref{varlt-conditioned} respectively. The result in  \eref{Gaussian-semi-infinite} is numerically verified in \fref{lt-semiinft-mean-l0}c.

%%%%%%%%%%%%%%%%%%%%%%%%%%%%%%%%%%%%%%%%%%%%%%%%%%%%%%%%%%
%%%%%%%%%%%%%%%%%%%%%%%%%%%%%%%%%%%%%%%%%%%%%%%%%%%%%%%%%

\section{Local time of diffusion with drift}
\label{drift-diffusion}
\noindent
In this section, we extend the problem of finding the distribution of the local time density to diffusion with a constant drift along the positive direction. We consider the problem with and without resetting. Here, as before, the particle resets to its initial position $x_0$ which is also the place where we compute the local time. Once again, we first look at the problem of finding the distribution of the local time density on an infinite domain. In the next section, we will consider the problem on a semi-infinite domain with an absorbing boundary at $x=0$.

\subsection{Infinite domain}
\label{potI}
%\subsubsection{$r=0$}
\noindent
As noticed before, to compute the moment generating function $\tilde{G}_r(x,x_0,k,\alpha)$ with resetting, one first needs to solve the problem without resetting (see Eq.~\eqref{renewal-formula}).
To do this, we first note that $Q_r(x_0,t)=1$ since there are no absorbing boundaries. Then, we solve 
Eq.~\eqref{GF-noReset-LT} with the following boundary conditions $\tilde{G}_0(|x| \to \infty,x_0,k,\alpha)=\frac{1}{\alpha}$ on the infinite line and get the following expression for the generating function 
\bea
\tilde{G}_{0}(x_0,x_0,k,\alpha)= \frac{1}{\alpha}
\frac{ \sqrt{\lambda^2+4D\alpha}}{k+\sqrt{\lambda^2+4D\alpha}}~,
\label{Gfr0potInf}
\eea
which is again independent of $x_0$ since the system is translation invariant. Substituting the above expression for $\tilde{G}_0$ in Eq.~\eqref{renewal-formula}, we obtain
\bea
\tilde{G}_{r}(x_0,x_0,k,\alpha)= \frac{\sqrt{\lambda^2+4D(r+\alpha)}}{\alpha \sqrt{\lambda^2+4D(r+\alpha)}+ k (r+\alpha)}~.
\eea

\begin{figure}[t]
\includegraphics[width=.45\hsize]{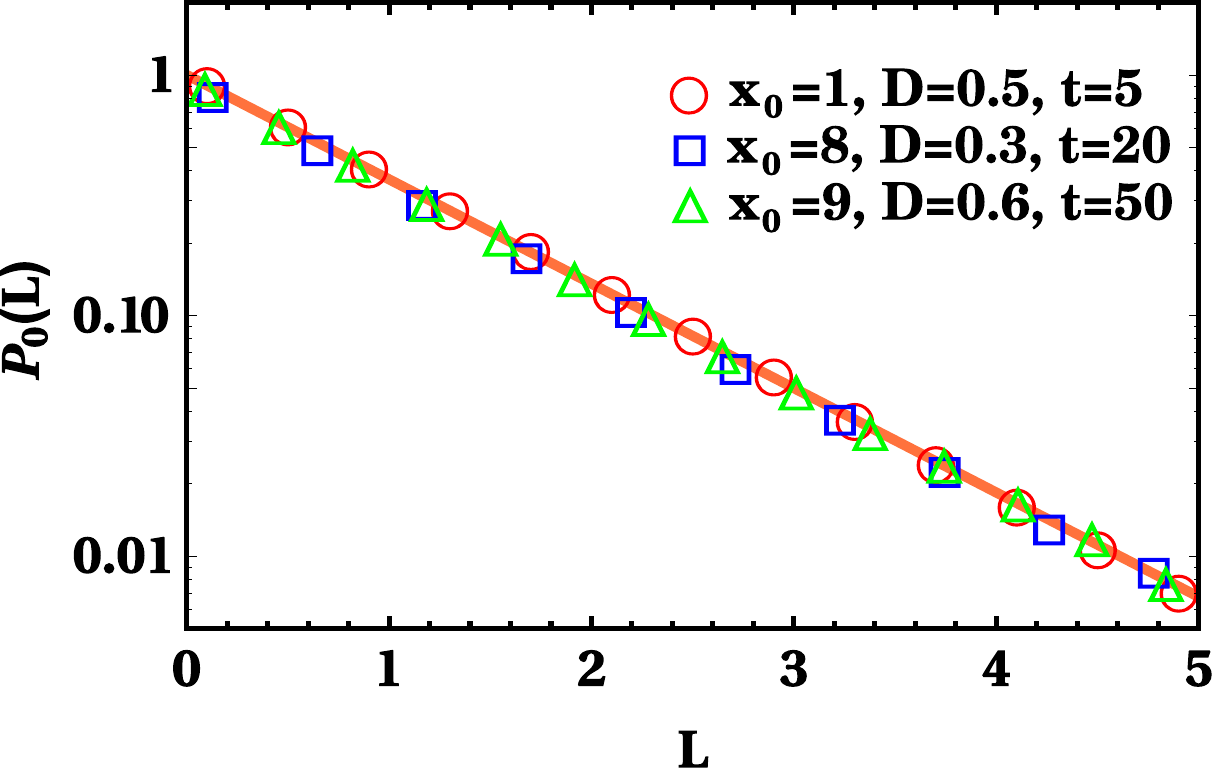}
\caption{The long time asymptotics of the probability distribution of the local time of a Brownian particle in the presence of drift ($\lambda=1$). Here, the solid orange line corresponds to the result from \eref{exact-recurrent} and results from numerical simulations are presented as symbols. Data collapse of the simulation results for $x_0=1,D=0.5,t=5$ (red circles); $x_0=8,D=0.3,t=20$ (blue squares) and $x_0=9,D=0.6,t=50$ (green triangles) verify that in the absence of resetting the probability distribution of the local time is asymptotically independent of the initial position $x_0$, the diffusion coefficient $D$, and the observation time $t.$}
\label{lt-inft-pdf-l1-r0}
\end{figure}

Following the contour integral procedure as discussed in the previous section one can, in principle, perform the inverse Laplace transforms with respect to $k$ and $\alpha$ and express the result in terms of an integral over a branch cut on the negative axis. However the expressions are not illuminating except for the $r=0$ case, where we find
\bea
P_0(L) &=& \frac{d J_\lambda(L,D|x_0,t)}{dL},~~~\text{where}  \\
J_\lambda(L,D|x_0,t)&=&1- e^{-\lambda L} - \frac{1}{2} \left[ e^{\lambda L}~\text{erfc}\left(L\sqrt{\frac{D}{t}} + \lambda \sqrt{\frac{t}{4D}} \right) - e^{-\lambda L}~\text{erfc}\left(-L\sqrt{\frac{D}{t}} + \lambda \sqrt{\frac{t}{4D}} \right) \right].
\eea
In particular, we observe that for large $t$ the above distribution becomes
\bea
P_0(L) \simeq \lambda~e^{-\lambda L}~,
\label{exact-recurrent}
\eea
which is independent of time to leading order. 
This happens because the particle drifts away from its initial position. As a result, trajectories hardly return to this position at the long time limit. Thus, most of the contribution to the local time density comes from initial excursions of the trajectories around $x_0$ and additional contributions to the local time become diminishingly small as time progresses. 
The result in \eref{exact-recurrent} is numerically verified in \fref{lt-inft-pdf-l1-r0}. It is worth pointing out that in \cite{Sabhapandit:2006}, the authors studied the local time density in a similar set up but in the presence of the potential $V(x)=\lambda|x|$. It was shown that the local time density, even in this case, attains an exponential form $P_0(L)=2\lambda e^{-2\lambda L}$ in the long time limit \cite{Sabhapandit:2006}.
%%%%%%%%%%%%%%%%%%%%%%%%%%%%%%%%%%%%%%%%%%%%%%%%%%%%%%%%%%
%%%%%%%%%%%%%%%%%%%%%%%%%%%%%%%%%%%%%%%%%%%%%%%%%%%%%%%%%%%

There is an alternative way to derive \eref{exact-recurrent} and explain why the distribution of the local time is exponential. For this we start by looking at a biased random walk on a one dimensional lattice of lattice constant $a$ and take the continuum limit at the end. Assume that the random walker starts at the origin and that at every time step $\Delta t$ it jumps to the right with probability $p=1/2+\epsilon$ and to the left with probability $q=1/2-\epsilon$. We now ask what is the probability that the particle returns to the origin $k$ times in $n$ steps. Let us denote this probability by $S(k,n)$. It can be shown that this probability can be expressed in terms of the probability $\mathbb{G}(x,n)$ to find the particle at position $x$ after $n$ steps. Following \cite{Hughes-book}, it is possible to show that the $z$-transform of 
$S(k,n)$, defined as $\tilde{S}(k,z)=\sum_{n=0}^\infty z^n~S(k,n)$, can be expressed in terms of 
$\tilde{g}(z)=\sum_{n=0}^\infty z^n~\mathbb{G}(0,n)$, as 
\bea
\tilde{S}(k,z)=\frac{1}{1-z}~\frac{1}{\tilde{g}(z)}~\left[1- \frac{1}{\tilde{g}(z)} \right]^k.
\eea
Starting from the master equation for $\mathbb{G}(x,n)$, one can show that $\tilde{g}(z)=\left[ 1-(1-4\epsilon^2)z^2\right]^{-1/2}$  (see \aref{generating-function} for a detailed derivation). Inserting this expression in the above equation for $\tilde{S}(k,z)$ and performing the inverse $z$-transform we obtain
\bea
S(k,n)|_{n \to \infty} = 2 \epsilon~e^{k \ln(1-2\epsilon)}~,
\eea
which is independent of $n$ in the large time limit. To make connection with the original problem of Brownian motion we need to go to the continuum limit by taking the drift $\epsilon$, the lattice constant $a$, and jump time duration $\Delta t$ to zero while keeping $D=a^2/\Delta t$ and $\lambda = 2 a \epsilon /\Delta t$ fixed (see \aref{generating-function} for more details). Identifying the local time density as $L=k \Delta t/a$ we find
\bea
S(k,n)|_{n \to \infty} \to  \lambda~e^{-\lambda L} dL~, 
\label{recurrent}
\eea
which is exactly the result we have obtained through the  generating function procedure earlier.

\begin{figure}[t]
\includegraphics[width=5cm,height=3.5cm]{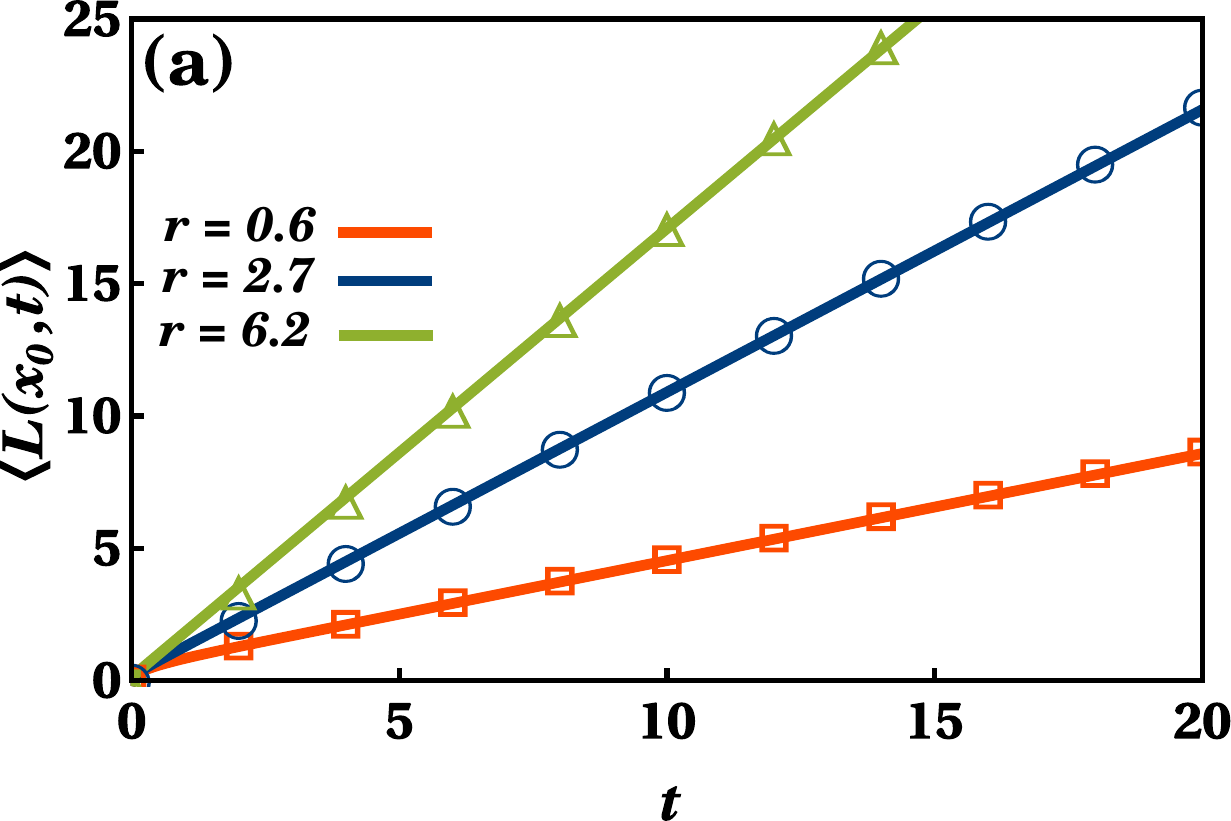}
\includegraphics[width=5cm,height=3.5cm]{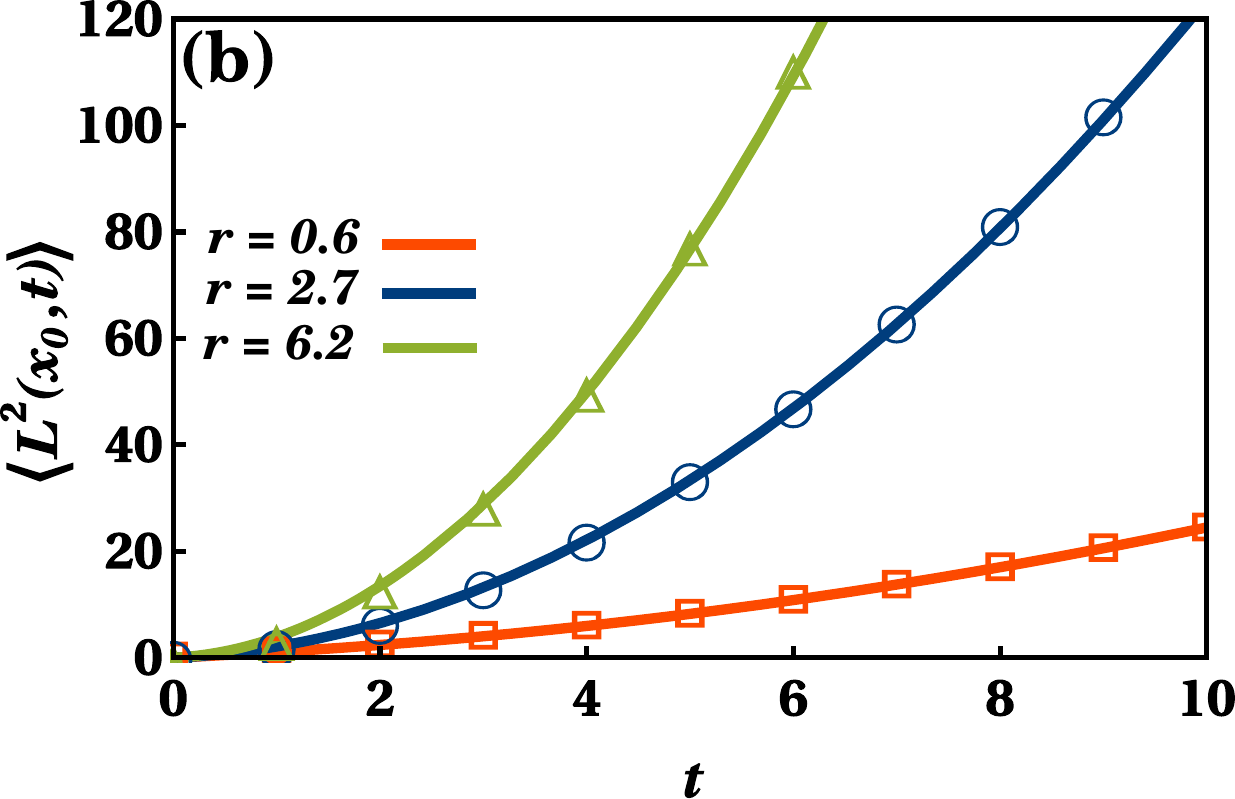}
\includegraphics[width=5cm,height=3.5cm]{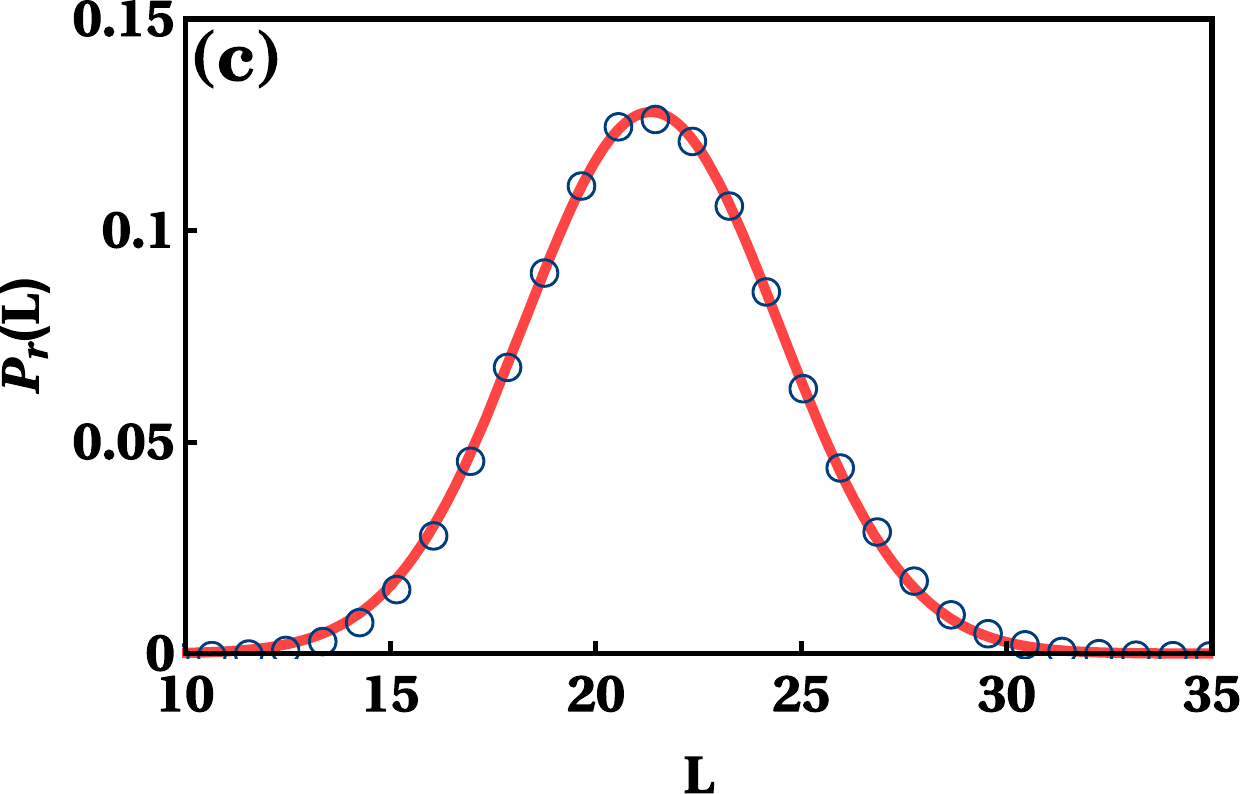}
\caption{The mean (a) and the second moment (b) of the local time of a Brownian particle in an infinite domain in the presence of drift ($\lambda=1).$ Here, $x_0=1,D=0.5$, and plots are made for three different values of the resetting rate: $r=0.6$ (orange), $2.7$ (blue) and $6.2$ (green). Solid lines come from Eqs. (\ref{mean-Infinite-pot}) and (\ref{sm-Infinite-pot}) and symbols come from numerical simulations. (c) The Gaussian approximation for the distribution of the local time (solid red line) and results from a numerical simulation (empty circles). Here,  $x_0=1,D=0.5,r=2.7$ and $t=20$.}
\label{lt-inft-mean-secmom-l1}
\end{figure}

Coming back to the case with resetting ($r>0$), we once again realize that in the large time limit the local time is a sum of multiple independent and identically distributed contributions that are gathered in between resetting events. The asymptotic distribution of $P_r(L)$ should hence be Gaussian and could thus be fully determined by its first two moments. These are given by
\bea
\langle L(x_0,t)  \rangle &=& \sqrt{\frac{4Dt}{\pi}}~\frac{r}{\lambda^2+4Dr}e^{-\frac{\lambda^2+4Dr}{4D}t}+ \frac{2Dr(1+2rt)+\lambda^2(1+rt)}{(\lambda^2+4Dr)^{3/2}}\erf \left[ \sqrt{\frac{(\lambda^2+4Dr)~t}{4D}} \right] ~,
\label{mean-Infinite-pot} \\
\text{and~~~}\langle L^2(x_0,t)  \rangle &=& \frac{16D^2 r^3 t(2+rt)+8D r^2 t (3+rt)\lambda^2}{(\lambda^2+4Dr)^3}+ \frac{2+rt(4+rt)-2e^{-\frac{\lambda^2+4Dr}{4D}t}}{(\lambda^2+4Dr)^3} \lambda^4,
\label{sm-Infinite-pot}
\eea
which in the large $t$ limit gives us
\bea
\langle L(x_0,t)  \rangle &\simeq& \frac{rt}{\sqrt{\lambda^2+4Dr}}~, \label{mean-Infinite-pot-large-t} \\
\sigma_r^2 &\simeq& 2 rt\frac{\lambda^2+2Dr}{(\lambda^2+4Dr)^2}+\frac{\lambda^4-4D^2 r^2-4Dr \lambda^2}{(\lambda^2+4Dr)^3}~.
\label{sm-Infinite-pot-large-t}
\eea  
The results in \eref{mean-Infinite-pot} and \eref{sm-Infinite-pot}, as well as the asymptotically Gaussian form of the local time distribution, are numerically verified in \fref{lt-inft-mean-secmom-l1}.

%%%%%%%%%%%%%%%%%%%%%%%%%%%%%%%%%%%%%%%%%%%%%%%%%%%%%%%%%%
%%%%%%%%%%%%%%%%%%%%%%%%%%%%%%%%%%%%%%%%%%%%%%%%%%%%%%%%%

\subsection{Semi-infinite domain with an absorbing boundary at the origin}
\begin{figure}[t]
\includegraphics[width=5cm,height=3.5cm]{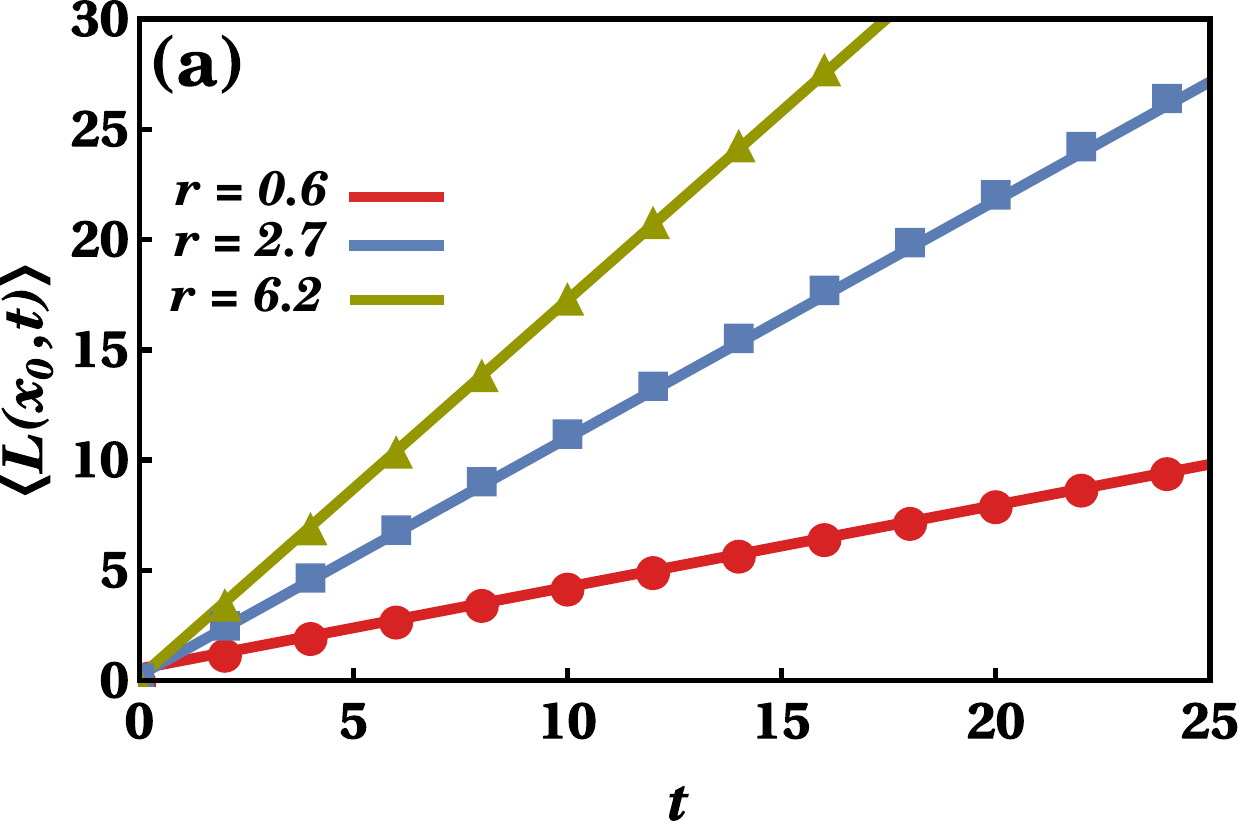}
\includegraphics[width=5cm,height=3.5cm]{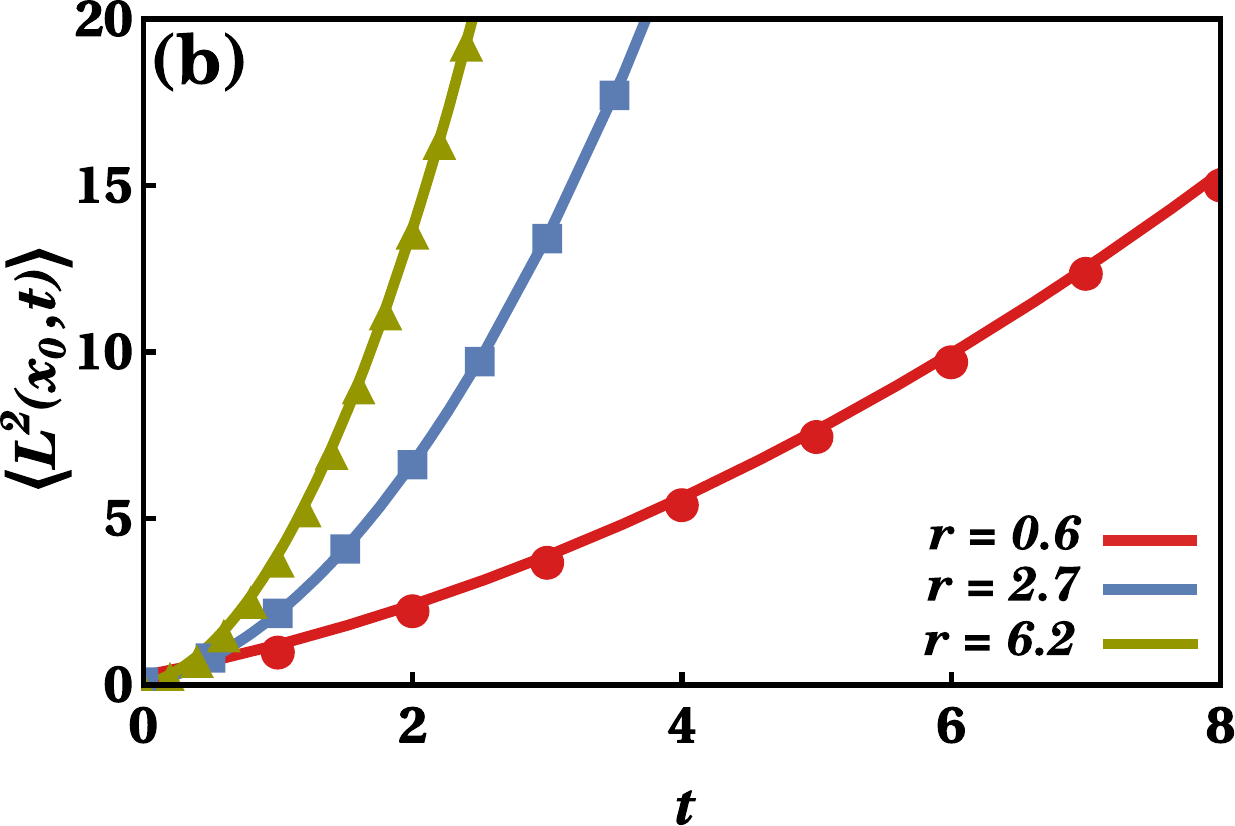}
\includegraphics[width=5cm,height=3.5cm]{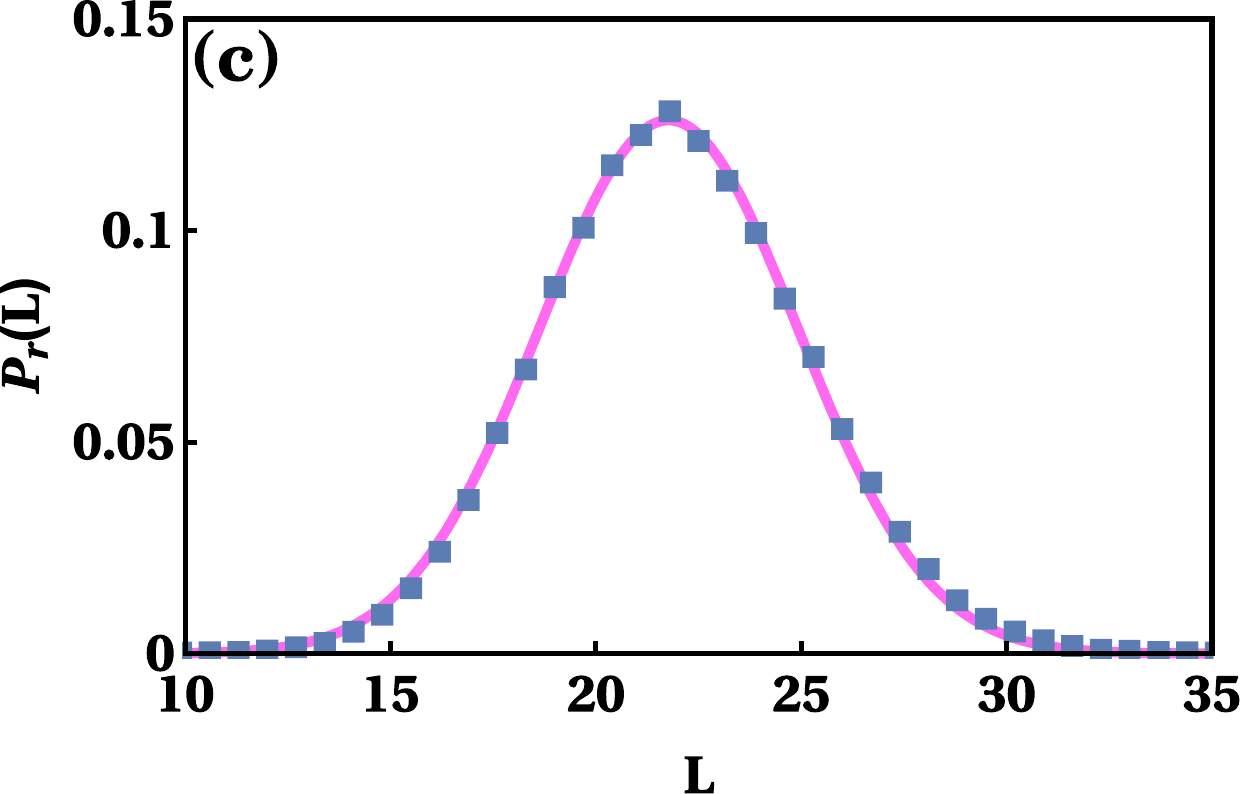}
\caption{The mean (a) and the second moment (b) of the local time of a Brownian particle in a semi-infinite domain in the presence of drift ($\lambda=1$). Solid lines come from Eqs. (\ref{meanltpot-conditioned}) and (\ref{smltpot-conditioned}) and symbols come from numerical simulations. Here, $x_0=1$, $D=0.5$, and plots were made for there different values of the resetting rate $r=0.6$ (red), $2.7$ (blue) and $6.2$ (green). (c) The Gaussian approximation for the distribution of the local time (solid magenta line) and results from a numerical simulation (full squares). Here, $x_0=1,D=0.5,r=2.7$ and $t=20$.}
\label{lt-semiinft-mean-lm1}
\end{figure}

\l{pot}
\noindent
In this section we study how the distributions obtained in section \ref{potI} get modified in the presence of an absorbing boundary at the origin $x=0$. Due to the presence of the drive away from the origin a particle which starts at $x_0$ will ultimately survive  with a finite probability $Q_0(x_0,t)|_{t \to \infty}=(1-e^{-\lambda x_0/D})$ \cite{Molini}. On the other hand, in presence of resetting, the particle is brought back to the resetting position $x_0$ intermittently. As a result, for any finite value of the resetting rate $r$ the survival probability decays exponentially with time (see \aref{survival-MFPT} and in particular \eref{qrSIpot} for more details)
\bea
Q_r(x_0,t) &\sim& e^{\alpha_0 t} ~\mathcal{J}(\alpha_0,b,Pe)~,\nonumber \\
\text{where~~~}\mathcal{J}(\alpha_0,b,Pe)&=&\frac{1-\exp \left[ -Pe(1+K_{r+\alpha_0})  \right]}{1-\frac{b^2}{2Pe K_{r+\alpha_0}} \exp \left[ -Pe(1+K_{r+\alpha_0})  \right]}~.
\label{Qr-asymptotic-pot}
\eea
Here, $\alpha_0$ 
is the solution of the transcendental equation $\alpha_0+r e^{-Pe(1+K_{r+\alpha_0})}=0$ with $K_\alpha=\sqrt{1+\frac{4D\alpha}{\lambda^2}}$ and $Pe=\frac{\lambda x_0}{2D}$ standing for the P\'eclet number. 
Here too, we find that in the long time limit the distribution of the local time density is Gaussian with 
\bea
\langle L_t(x_0) \rangle&\simeq& ~
\frac{g_1'(\alpha_0)+t g_1(\alpha_0)}{\mathcal{J}(\alpha_0,b,Pe)}~,
\l{meanltpot-conditioned} \\
\langle  L_t^2(x_0) \rangle&\simeq& ~\frac{g_2''(\alpha_0)+2 t g_2'(\alpha_0)+t^2 g_2(\alpha_0)}{\mathcal{J}(\alpha_0,b,Pe)}~,
\l{smltpot-conditioned} \\
\text{and~~~~~}\sigma^2_r(t)&\simeq& ~\frac{  \mathcal{J}(\alpha_0,b,Pe) g_2''(\alpha_0)-g_1'(\alpha_0)^2   +2t \left[ \mathcal{J}(\alpha_0,b,Pe) g_2'(\alpha_0)-g_1'(\alpha_0)g_1(\alpha_0) \right]  }{\mathcal{J}(\alpha_0,b,Pe)^2}~,
\l{varltpot-conditioned}
\eea
where the functions $g_1(\alpha)$ and $g_2(\alpha)$ are given by
\bea
g_1(\alpha)&=&\frac{\alpha+r}{\sqrt{\lambda^2+4D(\alpha+r)}}~\frac{\left( 1-e^{-2PeK_{r+\alpha}} \right) \left( 1-e^{-Pe \left(1+K_{r+\alpha}\right)} \right)}{\left[ 1-\frac{b^2}{2Pe K_{r+\alpha_0}}e^{-Pe(1+K_{r+\alpha_0})}+\frac{\alpha-\alpha_0}{2}e^{-Pe(1+K_{r+\alpha_0})}\frac{b^2D}{\lambda^2 K_{r+\alpha_0}^2} \left( 1+\frac{1}{Pe K_{r+\alpha_0}} \right)  \right]^2}~, \label{g1} \\
g_2(\alpha)&=& \frac{(\alpha+r)^2}{\lambda^2+4D(\alpha+r)}~\frac{\left( 1-e^{-2PeK_{r+\alpha}} \right)^2 \left( 1-e^{-Pe \left(1+K_{r+\alpha}\right)} \right)}{\left[ 1-\frac{b^2}{2Pe K_{r+\alpha_0}}e^{-Pe(1+K_{r+\alpha_0})}+\frac{\alpha-\alpha_0}{2}e^{-Pe(1+K_{r+\alpha_0})}\frac{b^2D}{\lambda^2 K_{r+\alpha_0}^2} \left( 1+\frac{1}{Pe K_{r+\alpha_0}} \right)  \right]^3}~,\label{g2}
\eea
where $\alpha_0$ here
is a solution of the transcendental equation $\alpha_0+r e^{-Pe(1+K_{r+\alpha_0})}=0$, with $K_{r+\alpha_0}=\sqrt{1+\frac{4D(r+\alpha_0)}{\lambda^2}}$ along with the other definitions $Pe=\lambda x_0/2D$, and $b=\sqrt{\frac{r}{D}}x_0$.
The results in \eref{meanltpot-conditioned} and \eref{smltpot-conditioned}, as well as the asymptotically Gaussian form of the local time distribution, are numerically verified in \fref{lt-semiinft-mean-lm1}. Note, however, that in comparison to Eqs.~\eqref{mean-Infinite-pot-large-t} and \eqref{sm-Infinite-pot-large-t} only the rates of growth get modified.
Hence, it is more interesting to look at the $r=0$ case in detail. In this case, following the same arguments given in the previous sections, one expects that in the large time limit the distribution of the local time will converge to a time independent exponential distribution. In what follows, we provide a derivation that verifies this expectation.

%%%%%%%%%%%%%%%%%%%%%%%%%%%%%%%%%%%%%%%%%%%%%%%%%%%%%%%%%%%
%%%%%%%%%%%%%%%%%%%%%%%%%%%%%%%%%%%%%%%%%%%%%%%%%%%%%%%%%

\begin{figure}[h!]
\includegraphics[width=.465\hsize]{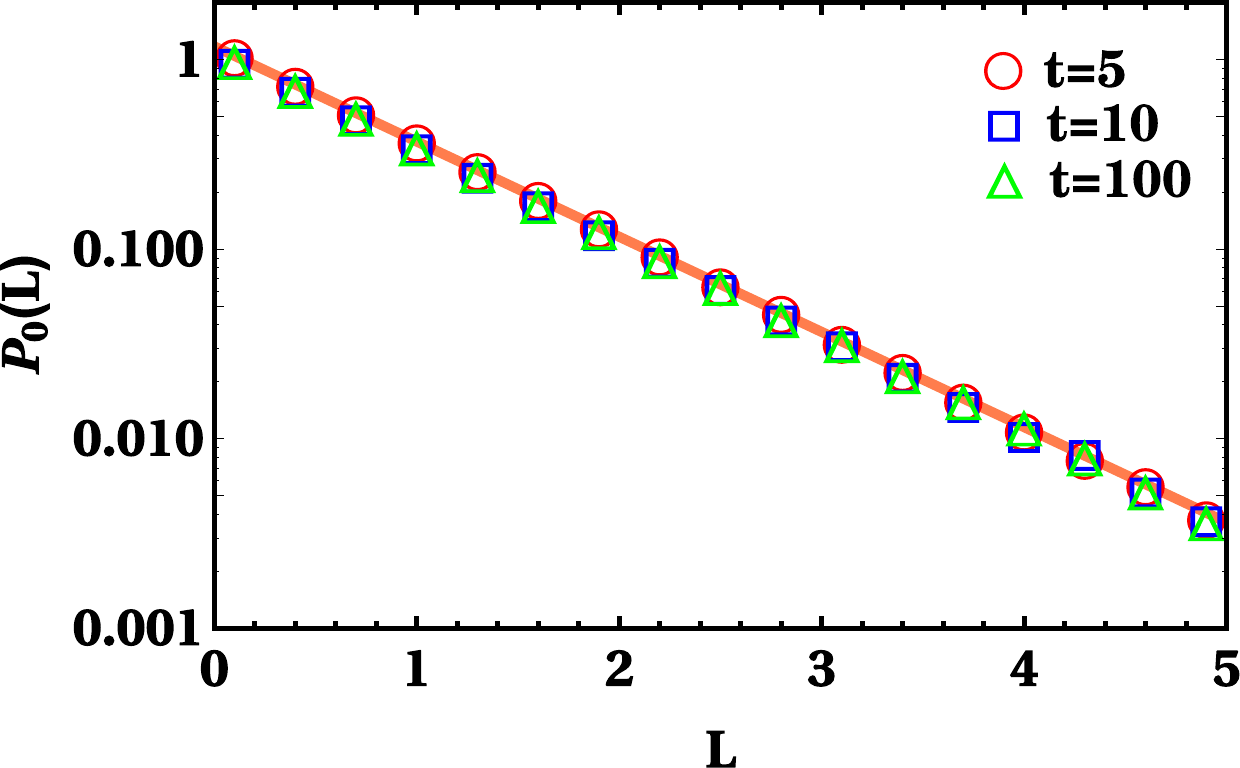}
\caption{The long time asymptotics of the probability distribution of the local time from \eref{PDFLT-SI-lambda-r0p0} (solid orange line) and results from simulations (symbols). Here, $x_0=1,$ $\lambda=1$ and $D=0.5$. Data collapse of the simulation results for $t=5$ (red circles), $t=10$ (blue squares) and $t=100$ (green triangles) verify that in the absence of resetting the probability distribution of the local time is asymptotically independent of the observation time $t.$}
\label{lt-semiinft-pdf-lm1-r0}
\end{figure}

We start by solving Eq.~\eqref{GF-noReset-LT} with $F(x)=\lambda >0$ and the following boundary conditions
$\tilde{G}_0(x \to \infty,x_0,k,\alpha) = \frac{1}{\alpha},~\tilde{G}_0(0,x_0,k,\alpha)=0$.
We obtain 
\bea
\tilde{G}_0(x_0,x_0,k,\alpha)=\frac{1}{\alpha} \left[ 1-\frac{\lambda K_{\alpha} e^{-Pe}+2k \sinh \left(Pe K_{\alpha} \right)}{\lambda K_{\alpha}e^{Pe K_{\alpha}}+2k \sinh \left( Pe K_{\alpha}\right)}  \right]~.
\label{GF-4}
\eea
Inverting \eref{GF-4} with respect to the conjugate variable $k$ we obtain
\bea
\tilde{F}_0(x_0,x_0,L,\alpha)= \frac{\lambda K_{\alpha} }{2 \alpha~\text{sinh} \left( Pe K_{\alpha}  \right) }~ \left( e^{Pe K_{\alpha}}-e^{-Pe} \right)~\exp \left[ -\frac{\lambda L K_{\alpha}}{2~\text{sinh} \left( Pe K_{\alpha}  \right)} e^{Pe K_{\alpha}} \right]~.
\eea
The large time asymptotics of the local time density function could be obtained by taking the $\alpha \to 0$ limit in $\tilde{F}_0(x_0,x_0,L,\alpha)$. In this limit, we have $K_\alpha \to 1$ and thus
\bea
\tilde{F}_0(x_0,x_0,L,\alpha \to 0) \sim \frac{\lambda}{\alpha} \exp \left[ -\lambda L e^{Pe } \text{cosech} \left( Pe   \right)/2 \right]~. 
\eea
Now performing the Laplace inversion with respect to $\alpha$, and utilizing the large time asymptotic form of the survival probability $Q_0(x_0,t) \simeq (1-e^{-2Pe})$, we
get the following asymptotic form for the local time density
\bea
P_0(L) = \frac{\lambda}{(1-e^{-2Pe})}  \exp \left[ -\frac{\lambda L}{(1-e^{-2Pe})} \right]~.
\label{PDFLT-SI-lambda-r0p0}
\eea
In \fref{lt-semiinft-pdf-lm1-r0}, we verify this result numerically for three large observation times.
%%%%%%%%%%%%%%%%%%%%%%%%%%%%%%%%%%%%%%%%%%%%%%%%%%%%%%%%%%
%%%%%%%%%%%%%%%%%%%%%%%%%%%%%%%%%%%%%%%%%%%%%%%%%%%%%%%%%

\section{Conclusions}
\l{conclusion}
\noindent
In this paper, we characterized the statistical properties of an additive Brownian functional --- the local time. We presented a general approach allowing for the computation of the local time of a Brownian particle that obeys overdamped Langevin dynamics and may also be subject to stochastic resetting. Then, we utilized this approach to compute the local time of such a particle in various situations. Namely, we analysed the local time with and without: resetting, an absorbing boundary, and drift. Out of these eight cases results were known only for the case where resetting, an absorbing boundary, and drift were all absent. Here, we focused on the remaining seven cases and presented a through analysis of each. Our findings can be summarized as follows. 

In the absence of resetting---but in the presence of an absorbing boundary, drift, or both---we find that the local time density tends to an asymptotic exponential form that is independent of the observation time. Specifically, when drift and an absorbing boundary at the origin are both present, we find
\bea
P_0(L) = \frac{\lambda}{(1-e^{-2Pe})}  \exp \left[ -\frac{\lambda L}{(1-e^{-2Pe})} \right]~,
\label{conc1}
\eea
as $t\to\infty$. Here, $\lambda=F(x)=-V'(x)$ stands for the constant drift and $Pe=\frac{\lambda x_0}{2D}$ is the P\'eclet number. To get some intuitive understanding of \eref{conc1} one should note that accumulation of local time can only come from consecutive visits of the particle to its initial position. However, having multiple such visits in the presence of drift is exponentially unlikely. Thus, even in the absence of an absorbing boundary, we find $P_0(L) = \lambda~e^{-\lambda L}$ (see \eref{exact-recurrent}), and note that this result can also be obtained by taking the limit $x_0\to\infty$ in \eref{conc1} which amounts to effective elimination of the absorbing boundary. Similarly, in the presence of an absorbing boundary, the particle runs the risk of being absorbed every time it returns to $x_0$ from far away. Thus, out of those trajectories that were not absorbed, finding a trajectory that visited $x_0$ a large number of times is once again exponentially unlikely. Consequently, even in the absence of drift, we find $P_0(L) = \frac{D}{x_0}e^{-\frac{D L}{x_0}}$ (see \eref{PDF-semi-Infinite-r0}), which can also be obtained directly by taking the limit $\lambda\to 0$ in \eref{conc1}. The compounded effect of drift and diffusion on the exponential decay rate, $\lambda/(1-e^{-2Pe})$, in \eref{conc1} is manifested by the appearance of the P\'eclet number. Indeed, note that as the drift increases, or as the diffusion coefficient decreases, the P\'eclet number increases and vice versa. In other words, low values of $Pe$ correspond to a diffusion-controlled regime, whereas high values of $Pe$ correspond to a drift-controlled regime. In that regard, one can trivially observe that in the limit $\lambda\to \infty$ particles never return to their initial position and \eref{conc1} gives a Dirac delta function at zero as expected.

In stark contrast to the above, we find that the distribution of the local time density in the presence of resetting is asymptotically Gaussian, i.e., admits  
\bea
P_r(L) \simeq \frac{1}{\sqrt{2 \pi \sigma^2_r(t)}}\text{exp}\left(-\frac{(L-\langle L(x_0,t)  \rangle)^2}{2 \sigma^2_r(t)} \right)~,
\label{conc2}
\eea
as $t\to\infty$, where the mean $\langle L(x_0,t) \rangle$ and variance $\sigma_r^2(t)$ of the local time density depend on the details of the problem. For example, in the absence of drift and an absorbing boundary we find $\langle L(x_0,t) \rangle=\frac{1}{2}\sqrt{\frac{r}{D}}t$ and $\sigma_r^2(t)=\frac{t}{4D}-\frac{1}{16r D}$. Asymptotically exact expressions for the mean and variance were also given in cases where an absorbing boundary and drift were present (see e.g., \eref{meanltpot-conditioned} and \eref{varltpot-conditioned}). These expressions are more involved, but it is important to note that we find that the mean and variance always grow linearly with the observation time. To intuitively understand this, as well as the Gaussian form in \eref{conc2}, it is enough to realize that in the presence of resetting---the total local time is a sum of many independent and identically distributed contributions that come from disjoint time intervals that stand in between resetting events. As the latter are typically $1/r$ units of time long the local time accumulates $\sim{rt}$ such  contributions over a long observation time $t$. Linear growth of the mean and variance follows immediately, and Gaussian asymptotics follows from the central limit theorem. 

This study was focused on  characterizing the local time statistics at $x_0$ --- which we took to be the initial and resetting position of the particle. However, the approach presented herein can also be extended to provide results for the local time at an arbitrary position $x$. Understanding how the moments and distribution of the local time depend on position, and in particular on the distance from $x_0$, would then be of prime interest. Going forward, it would also be interesting to study space-time correlations $\langle L(x,t)L(x',t') \rangle$ of the local time, and ultimately also its joint distribution, i.e., the probability that $L(x,t) = l$ and $L(x',t') = l'$. The local times of two adjacent points in space should be positively correlated. Indeed, one expects that finding one of these local times to be higher than average would increase the likelihood that the same could also be said for the other and vice versa. However, at this point it not clear how this correlation decays with the distance between the two points. Similarly, the local time is expected to display temporal correlations. If the local time at a given point in space is found to be higher than average it is likely to stay that way for some time. However, at this point it not clear how this correlation decays with the time separating two observations  and how it depends on the time at which the first observation was made. Answering these questions would significantly deepen our understanding of the local time and its behavior. 

%%%%%%%%%%%%%%%%%%%%%%%%%%%%%%%%%%%%%%%%%%%%%%%%%%%%%%%
\section{Acknowledgements}
Arnab Pal and Rakesh Chatterjee gratefully acknowledge support from the Raymond and Beverly Sackler Post-Doctoral Scholarship at Tel-Aviv University. Shlomi Reuveni acknowledges support from the Azrieli Foundation and from the Raymond and Beverly Sackler Center for Computational Molecular and Materials Science at Tel Aviv University. Anupam Kundu acknowledges support from DST grant under project No. ECR/2017/000634. This
work benefited from the support of the project 5604-2 of the Indo-French Centre for the Promotion of Advanced Research (IFCPAR).  We would also like to thank the Weizmann Institute
of Science for their hospitality during the SRitp workshop on correlations, fluctuations and anomalous transport in systems far from equilibrium as this work started there.

%%%%%%%%%%%%%%%%%%%%%%%%%%%%%%%%%%%%%%%%%%%%%%%%%%%%%%%

\appendix

\section{Survival probability and the mean first passage time}
\label{survival-MFPT}
In this paper, we have provided a comprehensive analysis of the local time statistics in the presence of an absorbing boundary for diffusion with and without drift.
Note that to compute the statistics in the presence of an absorbing boundary, it is important to have a prior knowledge of the survival probability. This, in turn, also serves as a natural platform to characterize the first passage properties of a Brownian particle in the presence of an absorbing boundary, drift, and resetting. In this appendix, we provide a detailed discussion on the survival probability of diffusion with stochastic resetting and explain how to compute them in various situations. We also explain how the mean first passage time to the boundary can be obtained directly from this analysis.

To compute the survival probability, it is always advantageous to use the backward Fokker Planck approach where one writes an evolution equation for the survival probability $Q_r(x,t)$ in terms of the initial variable. Utilizing the scheme presented in \cite{Restart1,Restart2}, we can write the following master equation for $Q_r(x,t)$ in the presence of a potential $V(x)$
\bea
\frac{\partial Q_r(x,t)}{\partial t}=D\frac{\partial^{2}Q_r(x,t)}{\partial x^{2}}+F(x)\frac{\partial Q_r(x,t)}{\partial x}-r[Q_r(x,t)-Q_r(x_0,t)]~.
\l{eq:S-eqn}
\eea
where recall $F(x)= - V'(x)$, and resetting takes place at the location $x_0$ \cite{Restart1}.
The Laplace transform $\tilde{Q}_r(x,\alpha)=\int_0^\infty dt~ e^{-\alpha t}Q_r(x,t)$ then satisfies
\bea
D\frac{\partial^{2}\tilde{Q}_r(x,\alpha)}{\partial x^{2}} + F(x)\frac{\partial \tilde{Q}_r(x,\alpha)}{\partial x}-(r+\alpha)\tilde{Q}_r(x,\alpha)=
-1 -r \tilde{Q}_r(x_0,\alpha)~.
\l{eq:s-tilde-eqn}
\eea
with the boundary conditions (i) $\tilde{Q}_r(0,\alpha)=0$, and (ii) $\tilde{Q}_r(x \to \infty,\alpha)$ is bounded from above by $1/\alpha$ \cite{Restart1,Restart2}.
First we recall that the result obtained for the survival probability in the absence of drift (i.e., $F(x)=0$), which is given by \cite{Restart1}
\bea
\tilde{Q}_r(x_0,\alpha)=\frac{1-e^{ - \sqrt{\frac{r+\alpha}{D}}~x_0 }}{\alpha+r e^{ - \sqrt{\frac{r+\alpha}{D}}~x_0 }}~.
\eea
Although in \cite{Restart1}, the authors inverted the expression for $\tilde{Q}_r(x_0,\alpha)$ to deduce the large $t$ asymptotic behavior of $Q_r(x_0,t)$, it was not known at all times. 
In \aref{ILT-derivation}, we explained how this can be done and provided an expression for $Q_r(x_0,t)$ that holds at all times (see \eref{surv-no-drift-w-reset}). 
From the survival probability one can immediately get the first passage time density $f_r(t)=-\frac{\partial Q_r(x,t)}{\partial t}$.
However, it is worth emphasizing that the mean first passage time can be computed directly from the Laplace transform $\tilde{Q}_r(x_0,\alpha)$ as well
\bea
\mathcal{T}_r(x_0)=\int_0^\infty~dt~ t f_r(t)=-\int_0^\infty~dt~ t\frac{\partial Q_r(x_0,t)}{\partial t}=\tilde{Q}_r(x_0,0)~.
\eea
This immediately results in
\bea
\mathcal{T}_r(x_0)=\frac{1}{r}\bigg [\exp\Big(\sqrt{\frac{r}{D}} x_0\Big)-1\bigg]~,
\l{mathcalT-r}
\eea
where $\sqrt{\frac{r}{D}}$ is an inverse length scale which measures the typical distance traveled by the particle between two resetting events \cite{Restart1}.
Interestingly, by carefully choosing an optimal resetting rate one can minimize the mean first passage time in \eref{mathcalT-r}. The condition for the minimum $d\mathcal{T}_r/dr|_{r=r^*}=0$ yields the following equation $
\frac{z^{*}}{2}=1-e^{-z^{*}},
%\label{optimize-flat-1}
$ where $z^*=\sqrt{\frac{r^*}{D}}x_0$.
This transcendental equation has a 
unique non-zero solution $z^{*}=1.593...$ from which one finds the optimal resetting rate $r^{*}=(z^*)^2D/x_0^2$ \cite{Restart1}. It is noteworthy that a similar analysis can also be made in the case of diffusion with drift.

To this end, we consider a particle that diffuses in
the presence of a drift $\lambda>0$ in the positive direction. As before, we solve \eref{eq:s-tilde-eqn} with the similar boundary conditions to obtain
the following solution for the Laplace transform $\tilde{Q}_r(x_0,\alpha)$
\bea
\tilde{Q}_r(x_0,\alpha)=\frac{1-e^{-Pe(1+K_{r+\alpha})}}{\alpha+r e^{-Pe(1+K_{r+\alpha})}}~,
\label{qrSIpot}
\eea
where we recall that $Pe=\lambda x_0/2D$ and $K_{\alpha}=\sqrt{1+\frac{2D \alpha}{\lambda^2}}$. It is possible to deduce the large $t$ asymptotic behavior of the survival probability using a complex inversion formula (as was done in \sref{semi-infinite-no-drift-rneq0}) and this is given by \eref{Qr-asymptotic-pot}. However, computation of
the mean first passage time turns out rather simple since $\mathcal{T}_r(x)=\tilde{Q}_r(x,0)$ which yields
\bea
\mathcal{T}_r(x_0)=\frac{1}{r}\bigg[\exp \Big(\frac{\lambda+\sqrt{\lambda^2+4Dr}}{2D}x_0 \Big )-1\bigg]~.
\l{sol:mfptx_0-unbound}
\eea

\begin{figure}[t]
\includegraphics[width=6.5cm]{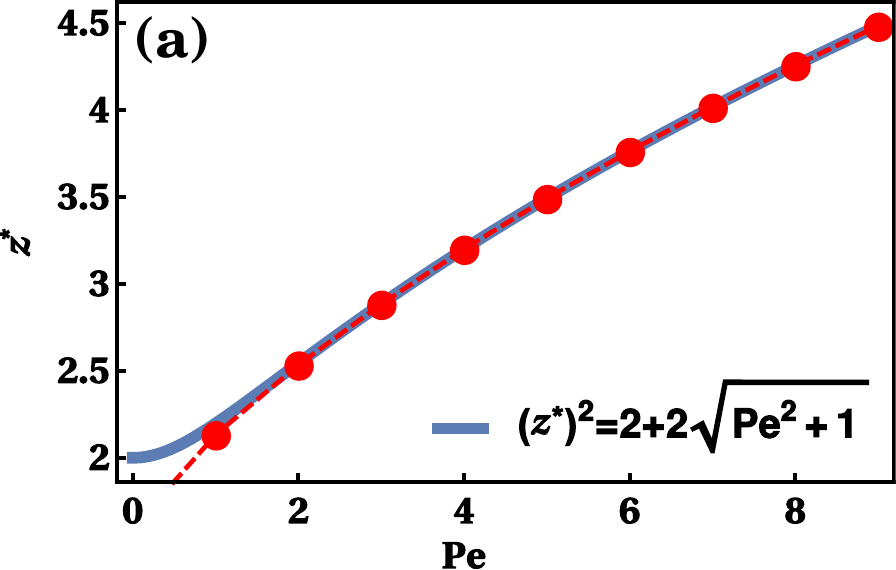}
\includegraphics[width=6.4cm]{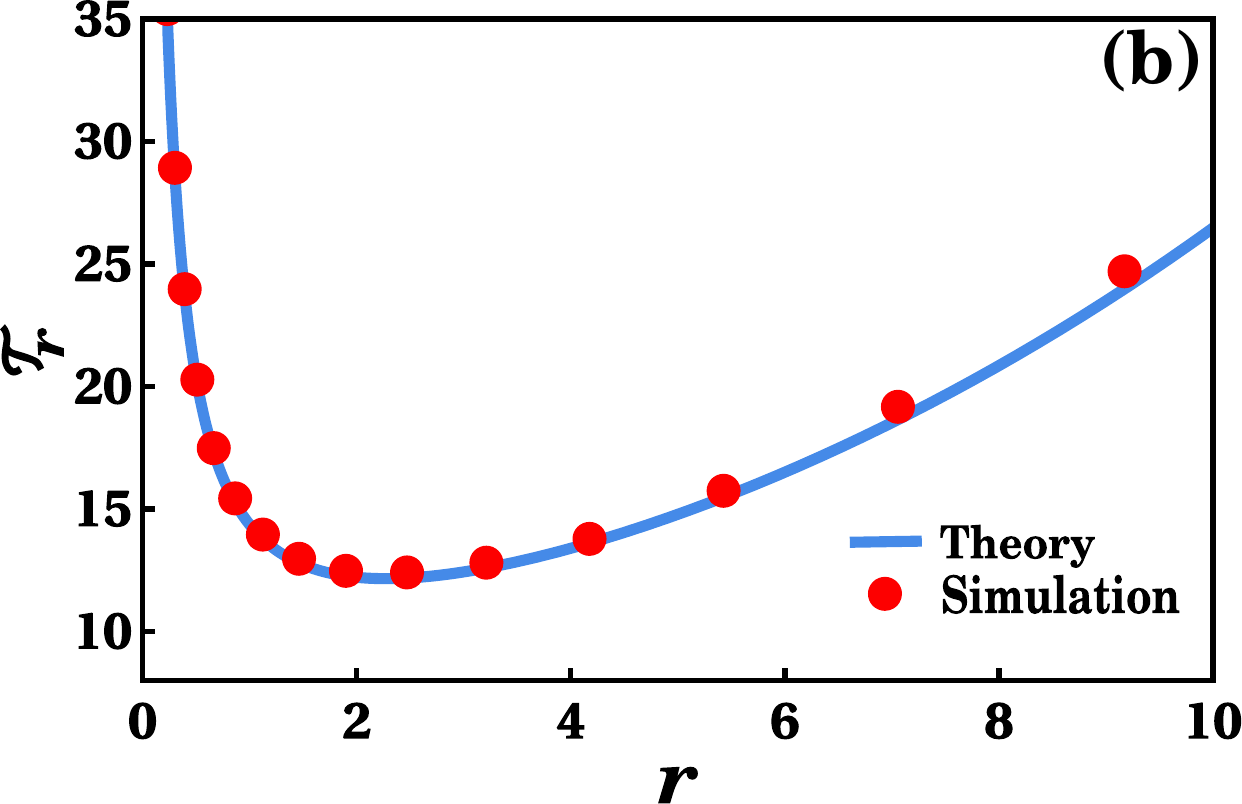}
\caption{(a) Variation of the root $z^*$ from \eref{optimize-unbounded-1} as a function of the P\'eclet number. The solid line corresponds to the asymptotic formula while the dashed line corresponds to the exact solution of $z^*$ as a function of $Pe$, as given by \eref{optimize-unbounded-1}.
Red circles indicate $z^*$ for different values of $Pe$ obtained from the simulations.
(b) Mean first passage time $\mathcal{T}_r$ of a Brownian particle in the presence of a drift $\lambda$ as a function of the resetting rate. We see an excellent match between the analytical result given by \eref{sol:mfptx_0-unbound} (solid blue line) and simulation results (full red circles). Here: $\lambda=1, ~x_0=1$ and $D=0.5.$}
\label{optimal-zstar}
\end{figure}

Note that $\underset{r \to 0}{\lim} \mathcal{T}_r(x_0) \sim \frac{1}{r}$ which diverges since the particle hardly experiences any resetting event, and in the absence of resetting the particle eventually drifts away from the origin. On the other hand, in the limit $r \to \infty$, the mean first passage time also diverges since the particle is localized around the resetting point. Evidently, this marks an existence for an optimal resetting rate which can be computed by setting $d\mathcal{T}_r/dr|_{r=r^*}=0$. Using the optimality condition in \eref{sol:mfptx_0-unbound}, we obtain the following equation for the optimal resetting rate
\bea
\frac{z^{*2}}{2\sqrt{Pe^2+z^{*2}}}=1-e^{- \left( Pe+\sqrt{Pe^2+z^{*2} } \right) }~,
\label{optimize-unbounded-1}
\eea
where recall $z^*=\sqrt{\frac{r^*}{D}}x_0$.
\eref{optimize-unbounded-1} asserts that the root $z^*$ depends on the P\'eclet number. While for small P\'eclet number, the limit is equivalent to the simple diffusion case, high P\'eclet number results in an approximate solution for $z^*$ given by
$z^{*2}=2+2\sqrt{Pe^2+1}$. This is evident from \fref{optimal-zstar}a. The solution for the optimal resetting rate can then easily be obtained by taking $r^*=Dz^{*2}/x_0^2$. We 
simulate this stochastic process and plot $\mathcal{T}_r$ as a function of the resetting rate against the analytical formula given by \eref{sol:mfptx_0-unbound}. We find an excellent agreement between the theory and the simulation (see \fref{optimal-zstar}b).

\section{Derivation of \eref{PDF-Infinite-r0p6} in the main text}
\label{LTInfinfite}
We start by recalling the auxiliary function in \eref{aux-1}
\bea
\tilde{F}_r(x_0,x_0,L,\alpha)=\frac{1}{a\sqrt{\alpha+r}}~\exp \left[ -\frac{\alpha}{a\sqrt{\alpha+r}}  T \right]~.
\eea
Let us first rewrite the above expression in the following manner
\bea
\tilde{F}_r(x_0,x_0,L,\alpha)
=\frac{1}{a\sqrt{\alpha+r}}~\exp \left[-\frac{T}{a}~\sqrt{\alpha+r}+\frac{r T}{a}~\frac{1}{\sqrt{\alpha+r}} \right]~.
\label{Fr-Laplace}
\eea
To proceed further, we first analyze the following function which has the form as the above function
\bea
\tilde{\mathcal{G}}(s+r)=\frac{1}{\sqrt{s+r}}~\exp[-m\sqrt{s+r}]~\exp[\frac{n}{\sqrt{s+r}}]~.
\eea
Therefore, Laplace inverse of the function
$\tilde{\mathcal{G}}(s+r)$ with respect to $s$ can be written using the properties of the Laplace transform
\bea
\mathcal{L}^{-1} \left[ \tilde{\mathcal{G}}(s+r) \right]=e^{-rt} \mathcal{G}(t)~,
\eea
which basically says it is enough to compute the Laplace inverse of the function
$\tilde{\mathcal{G}}(s)=\frac{1}{\sqrt{s}} \exp[-m\sqrt{s}]~\exp[\frac{n}{\sqrt{s}}]$, which we do in the following. In fact,
this can be seen as a product of two functions namely $\tilde{\mathcal{G}}_1(s)=\frac{1}{\sqrt{s}}\exp[-m\sqrt{s}]$
and $\tilde{\mathcal{G}}_2(s)=\exp[\frac{n}{\sqrt{s}}]$ so that the inverse Laplace transform of $\tilde{\mathcal{G}}(s)$ reads
\bea
\mathcal{G}(t)=\mathcal{L}^{-1} \left[ \tilde{\mathcal{G}}(s) \right] = \int_0^t~dt'~\mathcal{G}_1(t-t')~\mathcal{G}_2(t')~,
\label{G(t)}
\eea
where $\tilde{\mathcal{G}}_i(s)$ is the Laplace transform of $\mathcal{G}_i(t)$. We immediately note that $\tilde{\mathcal{G}}_1(s)$ has a standard Laplace inversion namely
\bea
\mathcal{G}_1(t)=\mathcal{L}^{-1} \left[ \tilde{\mathcal{G}}_1(s) \right]=\frac{1}{\sqrt{\pi t}}\exp \left[-\frac{m^2}{4t} \right]~.
\eea
On the other hand, inverse Laplace transform of $\tilde{\mathcal{G}}_2(s)$
can be written in terms of a complex inversion integral
\bea
\mathcal{G}_2(t)=\mathcal{L}^{-1} \left[ \tilde{\mathcal{G}}_2(s) \right]=\frac{1}{2\pi i}~\int_{\mathcal{C}-i\infty}^{\mathcal{C}+i\infty}~ds~\exp \left[ st \right]~\exp \left[ \frac{n}{\sqrt{s}} \right]~,
\eea
where $\mathcal{C}$ is the contour line along which one performs the integral. 
We now make a change of variables $s=n^2p$ and $\tau=n^2t$ so that 
the integral becomes
\bea
\mathcal{G}_2(t)=\frac{n^2}{2\pi i}~\int_{\mathcal{C}-i\infty}^{\mathcal{C}+i\infty}~dp~\exp \left[ p\tau \right]~\exp \left[ \frac{1}{\sqrt{p}} \right]~,
\eea
We can now expand the function $e^{\frac{1}{\sqrt{p}}}$ in an infinite  series so that the above integral admits the following form
\bea
\mathcal{G}_2(t) &=& n^2 \left[ \frac{1}{2\pi i}~\int~dp~\exp \left[ p\tau \right]~+ \sum_{j=1}^{\infty}~\frac{1}{2\pi i}~\int~dp~\frac{e^{p\tau}}{j!~p^{j/2}} \right] \nonumber \\
&=& n^2 \left[ \delta(\tau)~+ \sum_{j=1}^{\infty}~\frac{\tau^{\frac{j}{2} -1}}{j!~\Gamma(j/2)} \right]  \\
&=&  \delta(t)+ \sum_{j=1}^{\infty}~\frac{n^j t^{\frac{j}{2}-1}}{j!~\Gamma(j/2)} ~,
\eea
where we have used the fact $\delta(ax)=\delta(x)/|a|$.
Using this in \eref{G(t)}, we find
\bea
\mathcal{G}(t)&=&\int_0^t~dt'~\frac{\exp[-\frac{m^2}{4(t-t')}]}{\sqrt{\pi(t-t')}} \times  \left[ \delta(t')~+ \sum_{j=1}^{\infty}~\frac{n^{j}~t'^{\frac{j}{2}-1}}{j!~\Gamma(j/2)} \right] \\
&=& \frac{\exp[-\frac{m^2}{4t}]}{\sqrt{\pi t}}+\sum_{j=1}^{\infty}~\frac{n^j}{j!~\Gamma(j/2)}~\left[ \int_0^t~dt'~\frac{\exp[-\frac{m^2}{4(t-t')}] t'^{\frac{j}{2}-1}}{\sqrt{\pi(t-t')}} \right]~.
\eea
The integral in the last expression can be computed in the following way. Let us first define this integral as
\bea
H_j(\tilde{\alpha},t)=\frac{t^{\frac{j}{2}-1}}{\sqrt{\pi}}~\int_0^1~dz~\frac{\exp[-\frac{\tilde{\alpha}}{z}]}{\sqrt{z}}(1-z)^{j/2-1}~,
\eea
after taking $\tilde{\alpha}=m^2/4t$. This integration can be done exactly and we find
\bea
H_j(\tilde{\alpha},t)=t^{j/2-1}~\left[ \frac{\Gamma(j/2)}{\Gamma(\frac{1+j}{2})}~{}_{1}\mathcal{F}_{1} \left( \frac{1-j}{2},\frac{1}{2},-\tilde{\alpha} \right) -2 \sqrt{\tilde{\alpha}}~ {}_{1}\mathcal{F}_{1} \left( \frac{2-j}{2},\frac{3}{2},-\tilde{\alpha} \right)\right]~,
\eea
where $\Gamma(z)$ is the Gamma function.
Substituting this expression in \eref{G(t)}, we obtain
\bea
\mathcal{G}(t)=\frac{\exp[-\frac{m^2}{4t}]}{\sqrt{\pi t}}+\sum_{j=1}^{\infty}~\frac{n^j}{j!~\Gamma(j/2)}~H_j \left(\frac{m^2}{4t},t \right)~,
\eea
which is identical to \eref{PDF-Infinite-r0p6} in the main text. Therefore, Laplace inverse of the function
$\tilde{\mathcal{G}}(s+r)$ with respect to $s$ is given by
\bea
\mathcal{L}^{-1} \left[ \tilde{\mathcal{G}}(s+r) \right]=e^{-rt} \mathcal{G}(t)~.
\eea
Similarly, one can invert \eref{Fr-Laplace} with respect to the conjugate variable $\alpha$ by identifying 
$m$ and $n$ suitably.

%%%%%%%%%%%%%%%%%%%%%%%%%%%%%%%%%%%%%%%%%%%%%%%%%%%%%

\section{Derivation of \eref{surv-no-drift-w-reset}}
\label{ILT-derivation}

\begin{figure}[h]
\includegraphics[width=.3\hsize]{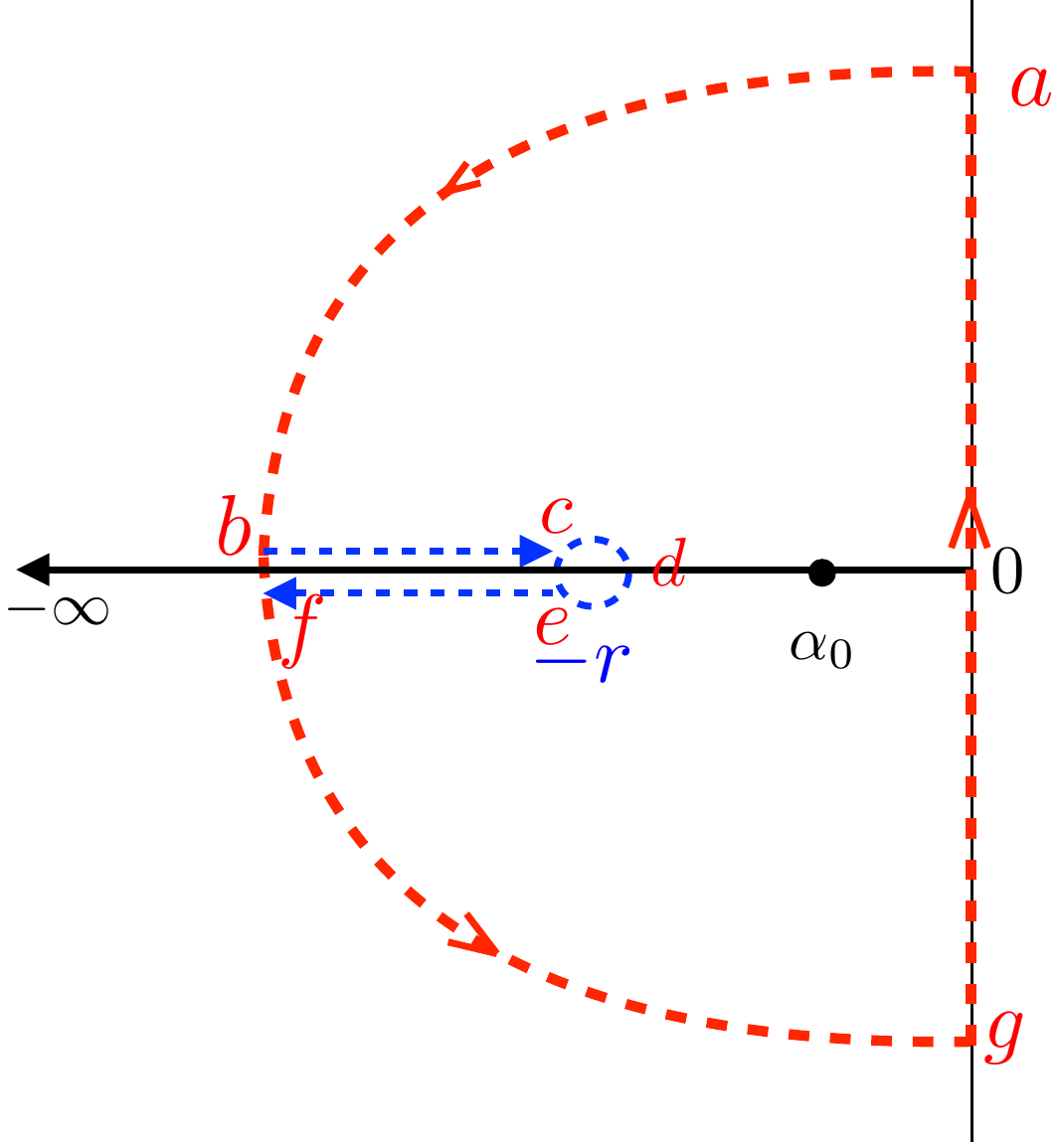}
\caption{The contour $\mathcal{B}$ used to perform the inverse Laplace transform of the survival probability in \eref{qrSI}.
There are two non-analytic points: A simple pole at $\alpha=\alpha_0 = -r(1- u_0)$ with $0 < u_0 < 1$ and a branch point at $\alpha=-r$.
In our calculation, we take the line $(-\infty,-r)$ along the real axis to be the branch cut.}
\label{contour_mt}
\end{figure}

We start by recalling the expression in \eref{qrSI} for the survival probability in Laplace space
\bea
\tilde{Q}_r(x_0,\alpha)=\frac{1-\exp \left[- \sqrt{\frac{r+\alpha}{D}}~x_0 \right]}{\alpha+r \exp \left[- \sqrt{\frac{r+\alpha}{D}}~x_0 \right]}~,
\label{qrSI-ex}
\eea
Following the complex inversion formula and the residue theorem, inverse Laplace transform of $\tilde{Q}_r(x_0,\alpha)$ can be formally written as 
\bea
Q_r(x_0,t)=\frac{1}{2\pi i} \int_{\mathcal{B}}~d\alpha~e^{\alpha t}~\tilde{Q}_r(x_0,\alpha)=\sum \text{~Residues~of~}e^{\alpha t}\tilde{Q}_r(x_0,\alpha)~\text{at~the~poles~of~}\tilde{Q}_r(x_0,\alpha)~,
\label{qrSI-ex-1}
\eea
where the integral is to be performed along a contour $\mathcal{B}$ as shown
in \fref{contour_mt}. 
Note that $\tilde{Q}_r(x_0,\alpha)$ in \eref{qrSI-ex} has a 
branch point at $\alpha=-r$ and a simple pole at $\alpha_0=-r(1-u_0)$ with $0 \le u_0 \leq 1$ satisfying
$u_0=1-\exp \left( -\sqrt{u_0} b\right)$ where $b=\sqrt{\frac{r}{D}}x_0$. This can be easily seen by setting the denominator of Eq.~\eqref{qrSI-ex} equal to zero i.e., $\alpha_0+r \exp \left[- \sqrt{\frac{r+\alpha_0}{D}}~x_0 \right]=0$. Note that the pole is in between $-r$ and $0$. 
Hence a natural choice for the branch cut is the line $[-\infty,-r]$ (see Fig.~\ref{contour_mt}). The integral along the contour can be decomposed in the following manner (see \fref{contour_mt})
\bea
\frac{1}{2\pi i} \int_{\mathcal{B}}~d\alpha~e^{\alpha t}~\tilde{Q}_r(x_0,\alpha)=  \frac{1}{2\pi i} \left[\int_{ga} +\int_{ab}+\int_{bc}+\int_{cde}+\int_{ef}+\int_{fg} \right] 
~d\alpha~e^{\alpha t}~\tilde{Q}_r(x_0,\alpha)
\label{contour-int}
\eea

Following the standard procedure of the complex inversion, one can immediately show that the integral along the chords `ab',~`cde', and `fg' vanish to zero and the contribution comes from the 
segments `bc' and `ef' along the branch cut which we evaluate in the following.
\bea
\int_{bc} d\alpha~e^{\alpha t}~\tilde{Q}_r(x_0,\alpha)&=&  \int_{-\infty}^{-r}d\alpha~e^{\alpha t} 
\frac{1-\exp \left[- \sqrt{\frac{r+\alpha}{D}}~x_0 \right]}{\alpha+r \exp \left[- \sqrt{\frac{r+\alpha}{D}}~x_0 \right]} \\
&=&- \int^{\infty}_{0}dz e^{-(r+z)t} \frac{1-\exp \left[ -ib_0\sqrt{z} \right] }{z+r-r\exp \left[ -ib_0\sqrt{z} \right]}
\label{bc}
\eea
where we have done the following change in variables $\alpha+r=ze^{\pi i}$ and introduced a new parameter $b_0=x_0/\sqrt{D}$.
Similarly, we can write
\bea
 \int_{ef} d\alpha~e^{\alpha t}~\tilde{Q}_r(x_0,\alpha)&=&
\int^{-\infty}_{-r}d\alpha~e^{\alpha t} 
\frac{1-\exp \left[- \sqrt{\frac{r+\alpha}{D}}~x_0 \right]}{\alpha+r \exp \left[- \sqrt{\frac{r+\alpha}{D}}~x_0 \right]} \\
&=&\int^{\infty}_{0}dz e^{-(r+z)t} \frac{1-\exp \left[ ib_0\sqrt{z} \right] }{z+r-r\exp \left[ ib_0\sqrt{z} \right]}
\label{ef}
\eea
where we have taken $\alpha+r=ze^{-\pi i}$ along the line `ef'.
Using the method of complex inversion formula, the integral along the line `ga' then yields the required integral and thus we find
\bea
Q_r(x_0,t)=\frac{1}{2\pi i} \int_{\mathcal{B}}~d\alpha~e^{\alpha t}~\tilde{Q}_r(x_0,\alpha)=\frac{1}{2\pi i}\int_{ga} d\alpha~e^{\alpha t}~\tilde{Q}_r(x_0,\alpha)
\eea
Now using \eref{contour-int} and \eref{qrSI-ex-1}, we can write
\bea
Q_r(x_0,t)=\frac{1}{2\pi i}\int_{ga} d\alpha~e^{\alpha t}~\tilde{Q}_r(x_0,\alpha)=\text{Residue~at~}\alpha_0-\frac{1}{2\pi i} \left[ \int_{bc} +\int_{ef} \right] d\alpha~e^{\alpha t}~\tilde{Q}_r(x_0,\alpha)
\label{contour-int-1}
\eea
Since $\alpha_0$ is a simple pole, we can immediately compute the residue which after simplification gives
\bea
\text{Residue~at~}\alpha_0=\lim_{\alpha \to \alpha_0}(\alpha-\alpha_0)e^{\alpha t}~\tilde{Q}_r(x_0,\alpha) =\frac{(1-e^{-b\sqrt{u_0}})}{1-(b/2\sqrt{u_0})~e^{-b\sqrt{u_0}}}e^{-r u_0 t}
\eea
Using the above result and substituting \eref{bc} and \eref{ef} in \eref{contour-int-1}, we finally arrive at
\bea
Q_r(x_0,t)=\frac{(1-e^{-b\sqrt{u_0}})}{1-(b/2\sqrt{u_0})~e^{-b\sqrt{u_0}}}e^{-r u_0 t}+h_r(b,t)
\eea
where 
\bea
h_r(b,t)&=&\frac{1}{2\pi i}\int_0^\infty dz ~e^{-(r+z)t}~\frac{2iz\sin(b_0\sqrt{z})}{(z+r)^2+r^2-2r(z+r)\cos(b_0\sqrt{z})} \\
&=&\frac{2e^{-rt}}{\pi}~\int_0^\infty dv~e^{-rtv^2}~\frac{v^3\sin(bv)}{v^4+2+2(1+v^2)[1-\cos(b v)]}~.
\eea
where to obtain the second line from the first in the above equation, we have made the following change of variable: $z=rv^2$. This concludes the derivation of \eref{surv-no-drift-w-reset}.

%%%%%%%%%%%%%%%%%%%%%%%%%%%%%%%%%%%%%%%%%%%%%%%%%%%%%%

\section{Detailed discussion on the probabilistic derivation of \eref{recurrent}}
\label{generating-function}
Consider the motion of a random walker on a one dimensional lattice. The walker moves to the right with probability $p_1$, and to the left with probability $p_2$. For this motion,
let us first define the following probabilities
\begin{itemize}
    \item  $\mathbb{G}(x,n)$: probability that a random walker is at $x$ after $n$-steps, given that it had started from the origin in 1-dimension. In particular, we define the probability density at the origin to be $g(n)\equiv \mathbb{G}(0,n)$. Since any return to the origin can occur only in even steps, $g(n)$ is defined only when $n$ is even.
    \item $f(n)$: probability that the walker returns to the origin at $n$-th step for the first time, given that it had started from the origin. This is the first passage probability of the walker to the origin.
    \item $q(n)$: probability that the walker never returned to the origin till the $n$-th step, given that it had started from the origin. This is the survival probability of the walker till $n$-th step.
    \item $S(k,n)$: probability that the walker came back to the origin $k$-times in $n$-steps, given that it had started from the origin.
\end{itemize}
Therefore, by definition, one can write
\bea
S(k,n)=\sum_{m_1}\cdots \sum_{m_k}\sum_{m'}~f(m_1) \cdots f(m_k)~q(m')~\delta_{n,\sum_{i=1}^{k}m_i+m'}~.
\label{recursion-1}
\eea
It will be convenient to define the $z$-transforms in the following way
\bea
\tilde{S}(k,z)&=&\sum_{m=0}^{\infty}~S(k,m)~z^m~, \\
\tilde{f}(z)&=&\sum_{m=0}^{\infty}~f(m)~z^m~, \\
\tilde{g}(z)&=&\sum_{m=0}^{\infty}~g(m)~z^m~.
\label{gf-1}
\eea
Now using the definition (\eref{gf-1}) in \eref{recursion-1}, we obtain
\bea
\tilde{S}(k,z)=\left[ \tilde{f}(z) \right]^k ~\tilde{q}(z)~.
\label{gf-2}
\eea
On the other hand, the propagator satisfies a recursive relation \cite{Hughes-book}
\bea
g(2n)=\delta_{n.0}+f(2n)+\sum_{m_1} \sum_{m_2}f(2m_1)f(2m_2)\delta_{n,m_1+m_2}+\sum_{m_1} \sum_{m_2}\sum_{m_3}f(2m_1)f(2m_2)f(2m_3)\delta_{n,m_1+m_2+m_3}+\cdots
\eea
from which, after performing a $z$-transformation, one can write $\tilde{g}(z)=\frac{1}{1-\tilde{f}(z)}$.
Also, the first passage probability $f(n)$ and the survival probability $q(n)$ are related to each other
by the relation $f(n)=q(n-1)-q(n)$. This implies that the generating functions satisfy this simple relation: $\tilde{q}(z)=\frac{1-\tilde{f}(z)}{1-z}$.  Using this relation in \eref{gf-2}, one arrives at the following relation between $S(k,n)$ and $g(n)$ in $z$-space
\bea
\tilde{S}(k,z)=\frac{1}{1-z}~\frac{1}{\tilde{g}(z)}~\left[1- \frac{1}{\tilde{g}(z)} \right]^k.
\l{gf-3}
\eea
We now
assign the following probabilities to the random walker: $p_1=1/2+\epsilon$, and $p_2=1/2-\epsilon$. Utilizing this, one can write the following master equation for the propagator 
\bea
\mathbb{G}(x,n)=\frac{1}{2} \left[ \mathbb{G}(x-1,n-1)+\mathbb{G}(x+1,n-1)  \right]+\epsilon \left[ \mathbb{G}(x-1,n-1)-\mathbb{G}(x+1,n-1)  \right]~.
\label{propagator-1}
\eea
The Fourier transform $\tilde{\mathbb{G}}(p,n)\equiv\sum_x~e^{ipx}\mathbb{G}(x,n)$ then satisfies
$\tilde{\mathbb{G}}(p,n)=(\cos{p}+2i \epsilon \sin{p} )\tilde{\mathbb{G}}(p,n-1)$, with the initial condition $\tilde{\mathbb{G}}(p,0)=\sum_x~e^{ipx}\mathbb{G}(x,0)=\sum_x~e^{ipx}\delta_{x,0}=1$. 
One can apply the method of iterations to find the solution to be
$\tilde{\mathbb{G}}(p,n)=(\cos{p}+2i \epsilon \sin{p})^n$. To compute $\mathbb{G}(x,n)$, one now needs to invert $\tilde{\mathbb{G}}(p,n)$ which gives us ${\mathbb{G}}(x,n)=\frac{1}{2\pi}\int_{-\pi}^{\pi}~dp~e^{-ixp}~(\cos{p}+2i \epsilon \sin{p})^n$. In particular,
the $z$-transform $\tilde{g}(z)=\sum_{n=0}^{\infty}~\mathbb{G}(0,n)z^n$ can be written in an integral form
\bea
\tilde{g}(z)&=&\frac{1}{2\pi} \int_{-\pi}^{\pi}~\frac{ dp}{1-(\cos{p}+2i \epsilon \sin{p})z} \\
&=&\frac{1}{2\pi} \int_{-\pi}^{\pi}dp~\int_0^\infty~dt~\exp \big(-\left[ 1-(\cos{p}+2i \epsilon \sin{p})z \right]t \big)\\
&=&\int_0^\infty~dt~ e^{-t} \frac{1}{2\pi} \int_{-\pi}^{\pi}dp~e^{zt(\cos{p}+2i \epsilon \sin{p})} \label{p-integral}\\
&=&\int_0^\infty~dt~ e^{-t} \mathcal{J}_0(zt\sqrt{1-4\epsilon^2}) \\
&=&\frac{1}{\sqrt{1-(1-4\epsilon^2)z^2}}~,
\l{g-z}
\eea
where $\mathcal{J}_0(x)$ is the Bessel function of first kind \cite{Ryzhik,Stegun}. To compute the $p$-integral in \eref{p-integral}, we introduce a new set of parameters ($A, B$) such that $A+B=zt,~A-B=2\epsilon zt$, and then identify the $p$-integral in \eref{p-integral} as a standard integral. This is given by
$\frac{1}{2\pi}\int_{-\pi}^{\pi}~dp~ e^{(A+B)\cos{p}+i(A-B)\sin{p}}=\mathcal{J}_0(2\sqrt{AB})$.
Next, we substitute $\tilde{g}(z)$ from \eref{g-z} in \eref{gf-3} to obtain an expression for $\tilde{S}(k,z)$. To derive back $S(k,n)$, we now use the Cauchy inversion formula which states
\bea
S(k,n)=\frac{1}{2\pi i}~\oint~dz~\frac{\tilde{S}(k,z)}{z^{n+1}}~,
\eea
where the closed integration contour encircles the origin and does not include any singularities of $\tilde{S}(k,z)$. By doing a change of variable $z=e^{-s}$, one can convert this integral into the standard inverse Laplace transform such that
\bea
S(k,n) \simeq \frac{1}{2\pi i} ~\int_{-i \infty}^{i \infty}~
ds~ e^{sn}~\frac{\sqrt{1-(1-4\epsilon^2)e^{-2s}}}{1-e^{-s}}~\left[ 1-\sqrt{1-(1-4\epsilon^2)e^{-2s}}  \right]^k~.
\eea
In the large $n$ limit, we find
\bea
S(k,n)_{n \to \infty} &\simeq& \frac{1}{2\pi i} \int_{-i \infty}^{i \infty}~\frac{2 \epsilon}{s} \left( 1-2\epsilon \right)^k \\
&\simeq& 2 \epsilon e^{k \ln(1-2 \epsilon)} \simeq 2 \epsilon e^{-2 \epsilon k}~,
\label{S-1}
\eea
which turns out to be independent of $n$.
Now to make connection with our original problem on Brownian motion we need to go to the continuum limit by taking the drift $\epsilon$, the lattice constant $a$ and the jump time duration $\Delta t$ to zero keeping $D=a^2/\Delta t$ and $\lambda = 2 a \epsilon /\Delta t$ fixed. Dividing both sides of \eref{propagator-1} by $\Delta t$ we find
\bea
\frac{\mathbb{G}(x,n)-\mathbb{G}(x,n-1)}{\Delta t}&=&\frac{a^2}{2\Delta t}  \frac{
\mathbb{G}(x-1,n-1)+\mathbb{G}(x+1,n-1)-2\mathbb{G}(x,n-1)}{a^2}  \nonumber \\
&+&\frac{a \epsilon}{\Delta t}~ \frac{ \mathbb{G}(x-1,n-1)-\mathbb{G}(x+1,n-1)}{a } ~.
\eea
Now we take the limits $a \to 0$, $\Delta t \to 0$ and $\epsilon \to 0$ while keeping $\lambda=2a\epsilon/\Delta t,~D=a^2/2\Delta t$ fixed such that 
 Fokker-Planck equation for the propagator $G(x,t)$ reads
\bea
\frac{\partial \mathbb{G}(x,t)}{\partial t}&=&\lim \limits_{\substack{a \to 0\\ \Delta t \to 0}}\left( \frac{a^2}{2\Delta t} \right) \partial_x^2 \mathbb{G}(x,t)-\lim \limits_{\substack{a,\epsilon \to 0\\ \Delta t \to 0}}\left(\frac{2 a \epsilon}{\Delta t} \right) \partial_x \mathbb{G}(x,t) \nonumber \\ 
&=& D\partial_x^2 \mathbb{G}(x,t)-\lambda \partial_x \mathbb{G}(x,t)~.
\eea
Note that this equation is exactly the Fokker-Planck equation for the probability density of a Brownian particle moving in one dimension in presence of a constant drive $\lambda$ in the positive $x$ direction with the correct identification of the parameters.
The local time density, on the other hand, can be identified as $L=k\Delta t/a$ in the continuum limit. Hence from \eref{S-1}, in the continuum limit we find,
\bea
S(k,n)_{n \to \infty} &\simeq& \lambda \frac{\Delta t}{a}e^{-\frac{\lambda \Delta t}{a}\frac{La}{\Delta t}} \simeq \lambda e^{-\lambda L} dL~,
\eea
which becomes independent of $k$. In other words, this implies that the distribution of the local time density is given by $P(L) \simeq \lambda e^{-\lambda L}$ as in \eref{exact-recurrent}, and independent of $t$.
%%%%%%%%%%%%%%%%%%%%%%%%=========================

%%%%%%%%%%%%%%%%%%%%%%%%%%%%%%%%%%%%%%%%%%%%%%%%%%%%%%%

\end{document}